\documentclass{IEEEtran}
\usepackage[T1]{fontenc}
\usepackage{amsmath,amssymb,amsfonts,amsthm}
\usepackage{graphicx}
\usepackage{textcomp,nicefrac}
\usepackage{bm}
\usepackage{microtype}

\DeclareFontShape{T1}{ptm}{m}{scit}{<->ssub * ptm/m/it}{}
\newtheorem{assumption}{Assumption}
\newtheorem{definition}{Definition}
\newtheorem{lemma}{Lemma}
\newtheorem{proposition}{Proposition}
\newtheorem{theorem}{Theorem}
\newtheorem{corollary}{Corollary}

\theoremstyle{remark}
\newtheorem{remark}{Remark}
\theoremstyle{plain}

\newcommand{\R}{\mathbb{R}}

\newcommand{\Nzero}{\mathbb{N}_0}
\newcommand{\Prob}{\mathbb{P}}
\newcommand{\Exp}{\mathbb{E}}
\newcommand{\ind}{\mathbf{1}}
\newcommand{\B}{\mathcal{B}}
\newcommand{\Id}{\operatorname{Id}}

\begin{document}
\setlength{\emergencystretch}{5em}
\setlength{\abovedisplayskip}{3pt}
\setlength{\belowdisplayskip}{3pt}
\setlength{\abovedisplayshortskip}{3pt}
\setlength{\belowdisplayshortskip}{3pt}

\title{Multi-Threshold Sampling: Local and Global Parameter Identifiability}

\author{Ao Qiu and Qingguo Xie
  \thanks{This work was supported in part by the National Natural Science Foundation of China under Grant 625B2079, Grant 61927801, Grant 62250002, and Grant 62050288. \textit{(Corresponding author: Qingguo Xie.)}}%
  \thanks{Ao Qiu is with the Department of Biomedical Engineering, Huazhong University of Science and Technology, Wuhan 430074 China (e-mail: aqiu@hust.edu.cn).}%
  \thanks{Qingguo Xie is with the Department of Biomedical Engineering, Huazhong University of Science and Technology, Wuhan 430074 China; with the Wuhan National Laboratory for Optoelectronics, Wuhan 430074 China; and with the Department of Electronic Engineering and Information Science, University of Science and Technology of China, Hefei 230026 China (e-mail: qgxie@hust.edu.cn).}}

\maketitle

\begin{abstract}
Multi-threshold (MT) sampling records threshold-crossing times for parameter estimation in applications including positron emission tomography, oil well logging, and photon-counting X-ray imaging. We study whether the distribution of recorded crossings uniquely determines the signal parameters and how many thresholds are required for this identifiability. For a fixed MT sampler, we show that the ranks of distribution derivatives with respect to the parameters cannot increase through crossing, timing, and selection. Under suitable regularity assumptions, we characterize local identifiability by a full-rank Jacobian of finitely many event probabilities defined on recorded crossings and establish a necessary and sufficient global identifiability criterion under additional rank and topological assumptions. For a nonhomogeneous Poisson signal model with four intensity parameters, a fixed known single-photon response, and deterministic MT sampling, we prove that four cumulative probabilities of the first recorded crossing time establish local and global identifiability under their respective conditions using a single sufficiently small positive threshold. This attains the minimum number of thresholds for the specified model and sampler. Numerical experiments illustrate these results and reveal weak parameter sensitivity and substantial finite-sample estimation errors when estimation uses these probabilities. The results establish parameter uniqueness while highlighting the role of statistical sensitivity in accurate estimation.
\end{abstract}

\section{Introduction}
\label{sec:introduction}

Multi-threshold (MT) sampling compares a signal with multiple thresholds and records the crossing times, typically using comparators and time-to-digital converters (TDCs)~\cite{xi2013fpga}. It represents the signal through crossing times, whereas conventional uniform time-domain sampling records amplitudes on a fixed time grid~\cite{xie2009potentials}. For high-speed signals, MT sampling can reduce data volume, hardware cost, and power consumption compared with high-rate uniform time-domain sampling~\cite{palka2017multichannel}. The recorded crossings can support parameter estimation and waveform reconstruction~\cite{kim2009multi}. Initially proposed for positron emission tomography (PET)~\cite{xie2005new}, MT sampling has nuclear science applications in time-of-flight PET~\cite{zhang2025performance}, oil well logging~\cite{zhao2023fpga}, and photon-counting X-ray imaging~\cite{zhang2024feasibility,chen2023multi}.

Parameter identifiability concerns whether the distribution of recorded crossings uniquely determines the signal parameters~\cite{rothenberg1971identification}. Local identifiability requires uniqueness near a given parameter value, while global identifiability requires uniqueness throughout the parameter set. Even when this uniqueness holds, estimation accuracy depends on parameter sensitivity, the observations used, and the amount of data~\cite{qiu2026prior}. Establishing identifiability therefore provides a starting point for studying the performance of MT estimators.

Our previous work developed a probabilistic framework that maps signal distributions to distributions of recorded crossings through crossing, timing, and selection~\cite{qiu2026sampling}. Building on this framework, we investigate whether these sampling stages preserve distinctions between parameter values. We address two questions: under what conditions are parameters locally or globally identifiable from recorded crossings, and how many thresholds suffice for identifiability under a specified signal model and MT sampler? A single threshold produces a distribution of crossing times whose probabilities may distinguish several parameters. The threshold requirement must therefore be established from the parameterized observation model.

The main contributions are as follows.

\emph{1) Local identifiability.} We show that the ranks of distribution derivatives with respect to the parameters cannot increase through crossing, timing, and selection. Under suitable differentiability and constant-rank assumptions, we characterize local identifiability by a nonsingular Jacobian of finitely many event probabilities defined on recorded crossings.

\emph{2) Global identifiability.} Under full-rank and topological assumptions, we prove that global identifiability is equivalent to properness of the parameter-to-distribution map onto its image: compact subsets of the distribution image have compact parameter preimages. This gives a necessary and sufficient criterion for global uniqueness under the stated assumptions.

\emph{3) Minimum number of thresholds.} For a nonhomogeneous Poisson signal model with four intensity parameters, a fixed known single-photon response, and deterministic MT sampling with time quantization, we prove that a single sufficiently small positive threshold ensures local and global identifiability under their respective conditions. The proofs use four cumulative probabilities of the first recorded crossing time, establishing that one threshold attains the minimum for the specified model and sampler.

Numerical experiments examine the sensitivity of these four probabilities and parameter estimation from repeated independent signal realizations sharing the same parameters. They illustrate the single-threshold results and quantify substantial finite-sample estimation errors associated with weak parameter sensitivity. These findings connect the identifiability analysis to the statistical behavior of the selected observations.

\section{Mathematical Preliminaries}
\label{sec:mathematical_preliminaries}

This section recalls the MT sampling framework of~\cite{qiu2026sampling} and specifies the observation model used in the identifiability analysis. We first define the signal and crossing spaces, then describe crossing, timing, and selection, and finally express the distribution of recorded crossings as the signal distribution acted on by a fixed sampling kernel. The same sampler and set of admissible signal paths are used for all parameter values.

\subsection{Signal and Crossing Spaces}
\label{subsec:signal_spaces}

Fix the observation window \(\mathcal T=[T_-,T_+]\), with \(T_-<T_+\). The signal path space \(\mathsf X=D(\mathcal T;\R)\) consists of real-valued paths right-continuous on \([T_-,T_+)\) with finite left limits on \((T_-,T_+]\). We use the uniform norm \(\|x\|_\infty=\sup_{t\in\mathcal T}|x(t)|\) to measure path magnitude and equip \(\mathsf X\) with the coordinate \(\sigma\)-algebra \(\B(\mathsf X)=\sigma\{x\mapsto x(t):t\in\mathcal T\}\). This is the Borel \(\sigma\)-algebra of the Polish Skorokhod \(J_1\) topology~\cite{billingsley1999convergence}. Subsets and finite products carry trace and product \(\sigma\)-algebras, respectively.

On a complete probability space \((\Omega,\mathcal F,\Prob)\), the random-signal space \(\mathfrak Y\) consists of measurable maps \(Y:\Omega\to\mathsf X\) satisfying \(\Exp\|Y\|_\infty<\infty\), identified when \(\Prob\{\|Y-Z\|_\infty=0\}=1\). We use measurable representatives when writing \(Y(t,\omega)=Y(\omega)(t)\) and identify deterministic paths with constant random elements.

For an integer \(N\geq1\), write \([N]=\{1,\ldots,N\}\). Crossing triples record time, threshold index, and direction mark in
\begin{equation}
  \mathsf E=\R\times[N]\times\{-1,-\tfrac12,0,\tfrac12,1\}.
  \label{eq:crossing_mark_space}
\end{equation}
Equip \(\mathsf E\) with its product Borel \(\sigma\)-algebra \(\B(\mathsf E)\). With \(\Nzero=\{0,1,\ldots\}\) and \(\delta_z\) denoting a Dirac measure, define the space of finite point measures by
\begin{equation}
  \mathsf N_{\mathrm f}(\mathsf E)
  =\left\{\sum_{i=1}^k\delta_{z_i}:k\in\Nzero,\ z_i\in\mathsf E\right\},
  \label{eq:finite_point_measure_space}
\end{equation}
whose \(\sigma\)-algebra, denoted by \(\B(\mathsf N_{\mathrm f}(\mathsf E))\), is generated by \(\nu\mapsto\nu(B)\), \(B\in\B(\mathsf E)\)~\cite{last2018lectures}. Here, \(k_\nu=\nu(\mathsf E)\) counts crossings with multiplicity; the empty sum is zero, and simple measures have unit atom masses.

\subsection{MT Sampling Framework}
\label{subsec:crossing_representation_operator}

The crossing, timing, and selection operators produce analog crossings, timed crossings, and recorded crossings, respectively.

The random threshold paths are \(\widetilde{\mathbf V}=\mathbf V+\mathbf b+\bm\eta\), where \(\mathbf V=(V_n)_{n=1}^N\in\R^N\) denotes nominal thresholds identified with constant paths, \(\mathbf b\in\mathsf X^N\) is deterministic threshold mismatch, and \(\bm\eta\) is measurable \(\mathsf X^N\)-valued threshold noise on a complete space \((\Omega_{\mathcal C},\mathcal F_{\mathcal C},\Prob_{\mathcal C})\). With \(\|\mathbf v\|_{\mathsf X^N}=\max_n\|v_n\|_\infty\), assume \(\Exp_{\mathcal C}\|\bm\eta\|_{\mathsf X^N}<\infty\) and \(\Exp_{\mathcal C}\eta_n(t)=0\) for all \(n,t\).

For \((x,\mathbf v)\in\mathsf X\times\mathsf X^N\), define the signal--threshold differences \(g_n(t;x,\mathbf v)=x(t)-v_n(t)\) and comparison states \(q_n(t;x,\mathbf v)=\operatorname{sgn}g_n(t;x,\mathbf v)\), with \(\operatorname{sgn}(0)=0\). Choose a Borel admissible domain \(\mathsf D_{\mathcal C}\subseteq\mathsf X\times\mathsf X^N\) such that, for every \((x,\mathbf v)\in\mathsf D_{\mathcal C}\) and \(n\in[N]\), there is a finite partition \(T_-=t_{n,0}<\cdots<t_{n,m_n}=T_+\) on whose open subintervals \(q_n(\cdot;x,\mathbf v)\) is constant. This condition ensures finitely many comparison-state changes and the existence of \(q_n(t^\pm)=\lim_{h\downarrow0}q_n(t\pm h)\) at interior times. Suppressing \((x,\mathbf v)\) on the right, define
\begin{equation}
  \begin{gathered}
    \mathcal R_n(x,\mathbf v)=\{t\in(T_-,T_+):\\
    |q_n(t^-)-q_n(t)|+|q_n(t)-q_n(t^+)|>0\},\\
    \sigma_n(t;x,\mathbf v)=\frac{q_n(t^+)-q_n(t^-)}{2}.
  \end{gathered}
  \label{eq:crossing_times_and_marks}
\end{equation}
Direction marks \(\pm1\), \(\pm\tfrac12\), and \(0\) indicate transitions between opposite sides, entry into or exit from equality, and isolated contacts, respectively; signs indicate increases or decreases in comparison state. Analog crossing times lie in the interior of the observation window. The crossing operator \(\mathcal C:\mathsf D_{\mathcal C}\to\mathsf N_{\mathrm f}(\mathsf E)\), defined by
\begin{equation}
  \mathcal C(x,\mathbf v)
  =\sum_{n=1}^N\sum_{t\in\mathcal R_n(x,\mathbf v)}
    \delta_{(t,n,\sigma_n(t;x,\mathbf v))},
  \label{eq:crossing_operator}
\end{equation}
produces simple finite point measures.

Timing and selection operators are measurable maps \(\mathcal Q,\mathcal H:\mathsf N_{\mathrm f}(\mathsf E)\to\mathsf N_{\mathrm f}(\mathsf E)\). Let \(\mathsf M=[N]\times\{-1,-\tfrac12,0,\tfrac12,1\}\) and \(\pi_{\mathrm m}(t,n,\sigma)=(n,\sigma)\). For \(\nu\in\mathsf N_{\mathrm f}(\mathsf E)\), the count measure on \(\mathsf M\) is \(((\pi_{\mathrm m})_{\#}\nu)(D)=\nu(\pi_{\mathrm m}^{-1}(D))\), \(D\subseteq\mathsf M\). Timing preserves this count measure:
\begin{equation}
  (\pi_{\mathrm m})_{\#}\mathcal Q(\nu)
  =(\pi_{\mathrm m})_{\#}\nu.
  \label{eq:deterministic_channel_mark_conservation}
\end{equation}
This is equivalent to preserving the crossing count for each threshold index and direction mark. Time coordinates may be continuous-valued or quantized; coincident triples retain multiplicity. Selection satisfies \(\mathcal H(\nu)\leq\nu\), where \(\mu\leq\nu\) means \(\mu(B)\leq\nu(B)\) for all Borel \(B\), and hence selects a submeasure of the timed crossings.

\emph{Measurability of the crossing operator.} Jointly measurable path evaluation and one-sided states give a Borel crossing relation with finite sections. On the standard Borel spaces \(\mathsf D_{\mathcal C}\) and \((T_-,T_+)\), the Lusin--Novikov theorem~\cite{kechris2012classical} provides a measurable enumeration of analog crossings. Counting these in each Borel subset of \(\mathsf E\) makes \(\mathcal C\) measurable.

\subsection{Distribution of Recorded Crossings}
\label{subsec:sampling_kernels_recorded_distributions}

We now represent the sampling stages by Markov kernels acting on distributions.

The common set of admissible signal paths
\begin{equation}
  \mathsf X_{\mathcal C}
  =\{x\in\mathsf X:\Prob_{\mathcal C}
     \{(x,\widetilde{\mathbf V})\in\mathsf D_{\mathcal C}\}=1\}
  \label{eq:admissible_signal_set}
\end{equation}
is Borel by measurability of integration. For \(x\in\mathsf X_{\mathcal C}\), define the stochastic crossing operator by \(\Phi_{\mathcal C}(x,\omega_{\mathcal C})=\mathcal C(x,\widetilde{\mathbf V}(\omega_{\mathcal C}))\) wherever \((x,\widetilde{\mathbf V}(\omega_{\mathcal C}))\in\mathsf D_{\mathcal C}\). This operator is jointly measurable on its domain and is defined \(\Prob_{\mathcal C}\)-almost surely for each admissible input \(x\). Probabilities involving partially defined sampling maps are evaluated on their probability-one domains.

For the stochastic timing and selection operators, indexed by \(\mathcal J\in\{\mathcal Q,\mathcal H\}\), take a complete auxiliary probability space \((\Omega_{\mathcal J},\mathcal F_{\mathcal J},\Prob_{\mathcal J})\) and a jointly measurable map \(\Phi_{\mathcal J}:\mathsf N_{\mathrm f}(\mathsf E)\times\Omega_{\mathcal J}\to\mathsf N_{\mathrm f}(\mathsf E)\). For \(\Prob_{\mathcal J}\)-almost every \(\omega_{\mathcal J}\), the section \(\Phi_{\mathcal J}(\cdot,\omega_{\mathcal J})\) is a timing or selection operator, respectively.

Assume that the internal randomness of the three stages is mutually independent, with product probability space \((\Omega_{\mathrm s},\mathcal F_{\mathrm s},\Prob_{\mathrm s})\) and \(\omega_{\mathrm s}=(\omega_{\mathcal C},\omega_{\mathcal Q},\omega_{\mathcal H})\). This independence concerns the internal randomness; stage outputs remain linked by sequential sampling. Write \(\Phi_{\mathcal J}^{\omega_{\mathcal J}}=\Phi_{\mathcal J}(\cdot,\omega_{\mathcal J})\) for \(\mathcal J\in\{\mathcal C,\mathcal Q,\mathcal H\}\).

A Markov kernel \(K\) assigns a probability measure \(K(a,\cdot)\) to each input \(a\), with \(a\mapsto K(a,B)\) measurable for every measurable output set \(B\). Its action on an input distribution and the composition of compatible kernels are
\begin{equation}
  \begin{aligned}
    (\mu K)(B)&=\int K(a,B)\,\mu(da),\\
    (K_1K_2)(a,B)&=\int K_2(b,B)\,K_1(a,db).
  \end{aligned}
  \label{eq:kernel_action_composition}
\end{equation}
The stage kernels are
\begin{equation}
  K_{\mathcal J}(a,B)
  =\Prob_{\mathcal J}\{\Phi_{\mathcal J}(a,\cdot)\in B\},
  \quad \mathcal J\in\{\mathcal C,\mathcal Q,\mathcal H\},
  \label{eq:stage_kernels}
\end{equation}
where \(a\in\mathsf X_{\mathcal C}\) for \(\mathcal J=\mathcal C\), \(a\in\mathsf N_{\mathrm f}(\mathsf E)\) for \(\mathcal J\in\{\mathcal Q,\mathcal H\}\), and \(B\in\B(\mathsf N_{\mathrm f}(\mathsf E))\). Joint measurability of the stochastic operators ensures the kernel property. The MT sampler \(\mathfrak S=(K_{\mathcal C},K_{\mathcal Q},K_{\mathcal H})\) has stochastic MT sampling operator
\begin{equation}
  \mathcal S(\cdot,\omega_{\mathrm s})
  =\Phi_{\mathcal H}^{\omega_{\mathcal H}}\circ
   \Phi_{\mathcal Q}^{\omega_{\mathcal Q}}\circ
   \Phi_{\mathcal C}^{\omega_{\mathcal C}}.
  \label{eq:stochastic_mt_sampling_operator}
\end{equation}
This operator is jointly measurable on the set of \((x,\omega_{\mathrm s})\in\mathsf X_{\mathcal C}\times\Omega_{\mathrm s}\) for which \((x,\widetilde{\mathbf V}(\omega_{\mathcal C}))\in\mathsf D_{\mathcal C}\).

Let \(M\geq1\) be an integer, let \(\Theta\subseteq\R^M\) be nonempty and Borel, and let \(\{Y_{\bm\theta}:\bm\theta\in\Theta\}\subseteq\mathfrak Y\) have distributions \(P_{Y_{\bm\theta}}\). Fix a sampler \(\mathfrak S\) common to all parameter values, with randomness independent of the signal, and assume \(P_{Y_{\bm\theta}}(\mathsf X_{\mathcal C})=1\) for every \(\bm\theta\). These conditions keep the sampling mechanism fixed as the parameter varies and ensure that all signal distributions are supported on the same set of admissible signal paths. In expressions involving sampling kernels, \(P_{Y_{\bm\theta}}\) denotes its restriction to \(\mathsf X_{\mathcal C}\).

On the signal--sampler product probability space, define the recorded crossings by \(\mathcal P_{\bm\theta}(\omega,\omega_{\mathrm s})=\mathcal S(Y_{\bm\theta}(\omega),\omega_{\mathrm s})\). Common admissibility and Fubini's theorem ensure that this composition is defined almost surely. Writing the MT sampling kernel as \(K_{\mathfrak S}(x,B)=\Prob_{\mathrm s}\{\mathcal S(x,\cdot)\in B\}\), independence and successive integration give
\begin{equation}
  K_{\mathfrak S}=K_{\mathcal C}K_{\mathcal Q}K_{\mathcal H},\qquad
  P_{\mathcal P_{\bm\theta}}=P_{Y_{\bm\theta}}K_{\mathfrak S}.
  \label{eq:parameterized_recorded_crossing_law}
\end{equation}
Here, \(\mathcal P_{\bm\theta}\) is a random finite point measure on \(\mathsf E\), whereas \(P_{\mathcal P_{\bm\theta}}\) is a probability measure on \(\mathsf N_{\mathrm f}(\mathsf E)\). The sampling kernel \(K_{\mathfrak S}\) is independent of \(\bm\theta\), so all parameter dependence in the distribution of recorded crossings enters through the signal distribution.

\section{Identifiability under MT Sampling}
\label{sec:identifiability}

This section develops local and global identifiability criteria from distributions of recorded crossings and selected event probabilities.

Using the model in Section~\ref{subsec:sampling_kernels_recorded_distributions}, define
\begin{equation}
  \Upsilon(\bm\theta)\triangleq P_{\mathcal P_{\bm\theta}}
  =P_{Y_{\bm\theta}}K_{\mathfrak S}.
  \label{eq:recorded_distribution_map}
\end{equation}
Throughout, \(\|\cdot\|_2\) denotes the Euclidean norm for vectors and the induced operator norm for linear maps between finite-dimensional spaces.

\begin{definition}[Local and global identifiability from distributions of recorded crossings]
\label{def:identifiability}
A parameter value \(\bm\theta_0\in\Theta\) is locally identifiable~\cite{rothenberg1971identification} if there exists \(r>0\) such that, for all \(\bm\theta\in\Theta\) with \(\lVert\bm\theta-\bm\theta_0\rVert_2<r\),
\begin{equation}
  P_{\mathcal P_{\bm\theta}}
  =P_{\mathcal P_{\bm\theta_0}}
  \quad\Longrightarrow\quad
  \bm\theta=\bm\theta_0.
  \label{eq:local_identifiability}
\end{equation}
The parameter is locally identifiable on \(\Theta\) if this holds at every \(\bm\theta_0\in\Theta\).

The parameter \(\bm\theta\) is globally identifiable on \(\Theta\) from recorded crossings under \(\mathfrak S\) if, for all \(\bm\theta,\bm\theta'\in\Theta\),
\begin{equation}
  P_{\mathcal P_{\bm\theta}}
  =P_{\mathcal P_{\bm\theta'}}
  \quad\Longrightarrow\quad
  \bm\theta=\bm\theta'.
  \label{eq:global_identifiability}
\end{equation}
\end{definition}

Local identifiability at \(\bm\theta_0\) concerns uniqueness of that parameter value among nearby alternatives. Injectivity of the distribution map on an entire neighborhood is a stronger property, obtained below under the full-rank condition.

An implementation may retain only sample means of selected recorded statistics. Event probabilities are expectations of indicator functions, so this also includes estimation from empirical event frequencies. We therefore consider whether the expectations of fixed recorded statistics uniquely determine the parameter.

\begin{definition}[Local and global mean identifiability with respect to recorded statistics]
\label{def:mean_identifiability}
Fix a measurable map \(\bm f:\mathsf N_{\mathrm f}(\mathsf E)\to\R^d\), with integer \(d\geq1\), common to all parameter values. Suppose \(\Exp\|\bm f(\mathcal P_{\bm\theta})\|_2<\infty\) for every \(\bm\theta\in\Theta\), and define
\begin{equation}
  \bm m_f(\bm\theta)\triangleq
  \Exp[\bm f(\mathcal P_{\bm\theta})]
  =\int\bm f(\nu)\,\Upsilon(\bm\theta)(d\nu).
  \label{eq:recorded_statistic_mean}
\end{equation}
A parameter value \(\bm\theta_0\in\Theta\) is locally mean identifiable with respect to \(\bm f\) under \(\mathfrak S\) if there exists \(r>0\) such that, for all \(\bm\theta\in\Theta\) with \(\|\bm\theta-\bm\theta_0\|_2<r\),
\begin{equation*}
  \bm m_f(\bm\theta)=\bm m_f(\bm\theta_0)
  \quad\Longrightarrow\quad\bm\theta=\bm\theta_0.
\end{equation*}
The parameter is locally mean identifiable on \(\Theta\) if this holds at every \(\bm\theta_0\in\Theta\), and globally mean identifiable on \(\Theta\) if \(\bm m_f\) is injective on \(\Theta\), in both cases with respect to the fixed \(\bm f\) under \(\mathfrak S\).
\end{definition}

\begin{remark}
Equality of distributions of recorded crossings implies equality of \(\bm m_f\), so local or global mean identifiability implies the corresponding identifiability in Definition~\ref{def:identifiability}; the converse need not hold for a fixed \(\bm f\). For each fixed \(\bm\theta\), let \((\mathcal P_{\bm\theta}^{(\ell)})_{\ell\geq1}\) be independent copies of \(\mathcal P_{\bm\theta}\). The integrability condition and the strong law of large numbers give
\begin{equation*}
  \frac1L\sum_{\ell=1}^L\bm f(\mathcal P_{\bm\theta}^{(\ell)})
  \xrightarrow[L\to\infty]{\mathrm{a.s.}}\bm m_f(\bm\theta).
\end{equation*}
\end{remark}

\begin{figure}[!ht]
  \centering
  \includegraphics[width=\columnwidth]{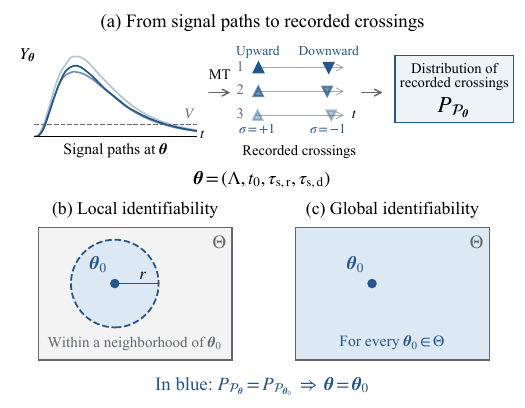}
  \caption{Local and global identifiability from recorded crossings. (a) Signal realizations at fixed \(\bm\theta\) are sampled by the same MT sampler. Each row of recorded crossings corresponds to the signal path of the same shade. The box denotes the distribution \(P_{\mathcal P_{\bm\theta}}\). (b) Local identifiability at \(\bm\theta_0\) means that its distribution uniquely determines \(\bm\theta_0\) within some neighborhood of radius \(r\). (c) Global identifiability requires this uniqueness throughout \(\Theta\) for every \(\bm\theta_0\in\Theta\). The parameter set and neighborhoods are schematic.}
  \label{fig:identifiability_definition}
\end{figure}

Figure~\ref{fig:identifiability_definition} illustrates Definition~\ref{def:identifiability} using the signal in Definition~\ref{def:poisson_photon_signal}.

\subsection{Parameter Derivatives under MT Sampling}
\label{subsec:parameter_derivatives}

A finite signed measure \(\zeta=\mu_1-\mu_2\) is a difference of finite nonnegative measures. Such measures on a fixed measurable space form a Banach space under the total variation (TV) norm
\begin{equation}
  \|\zeta\|_{\mathrm{TV}}
  =\sup_{\substack{g\text{ measurable}\\|g|\leq1}}
    \left|\int g\,d\zeta\right|.
  \label{eq:total_variation_norm}
\end{equation}
Kernel action extends linearly from \eqref{eq:kernel_action_composition}, with
\begin{equation}
  \|\zeta K\|_{\mathrm{TV}}\leq\|\zeta\|_{\mathrm{TV}}.
  \label{eq:kernel_tv_contraction}
\end{equation}
Indeed, \((Kg)(a)=\int g(b)K(a,db)\) is measurable with \(|Kg|\leq1\) whenever \(|g|\leq1\), and \(\int g\,d(\zeta K)=\int Kg\,d\zeta\). Thus kernel action is bounded and linear.

\begin{samepage}
\begin{assumption}[Differentiability of the signal distributions]
\label{ass:signal_distribution_differentiability}
The parameter set \(\Theta\subseteq\R^M\) is nonempty and open. The map \(\bm\theta\mapsto P_{Y_{\bm\theta}}\) is continuously Fr\'echet differentiable in the total variation norm.
\end{assumption}
\end{samepage}

Write \(D_{\bm\theta}\) for the Fr\'echet derivative with respect to \(\bm\theta\). For each direction \(\bm a\in\R^M\), the signed measure \(D_{\bm\theta}P_{Y_{\bm\theta}}[\bm a]\) is a TV limit of difference quotients concentrated on \(\mathsf X_{\mathcal C}\). Concentration on this set is preserved under TV limits, so these measures are fully determined by their restrictions to \(\mathsf X_{\mathcal C}\). For recorded crossings, the corresponding directional derivatives are signed measures on \(\mathsf N_{\mathrm f}(\mathsf E)\).

\begin{definition}[Ranks of parameter derivatives]
\label{def:parameter_derivatives}
Under Assumption~\ref{ass:signal_distribution_differentiability}, define the ranks of the distribution derivatives by
\begin{equation}
  \begin{aligned}
    r_Y(\bm\theta)
      &\triangleq\operatorname{rank}D_{\bm\theta}P_{Y_{\bm\theta}},\\
    r_{\mathcal C}(\bm\theta)
      &\triangleq\operatorname{rank}D_{\bm\theta}
        (P_{Y_{\bm\theta}}K_{\mathcal C}),\\
    r_{\mathcal Q}(\bm\theta)
      &\triangleq\operatorname{rank}D_{\bm\theta}
        (P_{Y_{\bm\theta}}K_{\mathcal C}K_{\mathcal Q}),\\
    r_{\mathcal H}(\bm\theta)
      &\triangleq\operatorname{rank}D_{\bm\theta}\Upsilon(\bm\theta).
  \end{aligned}
  \label{eq:parameter_derivative_ranks}
\end{equation}
Rank is the dimension of the derivative's range and counts linearly independent parameter derivatives.
\end{definition}

\begin{proposition}[Parameter derivatives through the sampling stages]
\label{prop:derivatives_through_sampling}
Under Assumption~\ref{ass:signal_distribution_differentiability}, for \(K\in\{K_{\mathcal C},K_{\mathcal C}K_{\mathcal Q},K_{\mathfrak S}\}\), the map \(\bm\theta\mapsto P_{Y_{\bm\theta}}K\) is continuously Fr\'echet differentiable in the total variation norm and satisfies
\begin{equation}
  D_{\bm\theta}(P_{Y_{\bm\theta}}K)[\bm a]
  =\bigl(D_{\bm\theta}P_{Y_{\bm\theta}}[\bm a]\bigr)K
  \label{eq:sampled_parameter_derivatives}
\end{equation}
for all \(\bm\theta\in\Theta\) and \(\bm a\in\R^M\). The ranks satisfy
\begin{equation}
  M\geq r_Y(\bm\theta)\geq r_{\mathcal C}(\bm\theta)
  \geq r_{\mathcal Q}(\bm\theta)\geq r_{\mathcal H}(\bm\theta).
  \label{eq:rank_through_sampling}
\end{equation}
\end{proposition}

\begin{samepage}
\begin{proof}
By \eqref{eq:kernel_tv_contraction}, each kernel acts as a bounded linear map on finite signed measures, so differentiation commutes with its action. The rank of a composition is bounded above by the rank of each factor, which gives \eqref{eq:rank_through_sampling}.
\end{proof}
\end{samepage}

\subsection{Conditions for Local Identifiability}
\label{subsec:local_identifiability}

The rank inequalities describe first-order parameter distinctions. A strict rank decrease identifies a sampling stage at which some previously distinguishable directions become indistinguishable to first order. The local criterion below connects these derivative ranks to parameter uniqueness.

The probability criterion uses bounded event indicators, whose expectations inherit the TV differentiability established above. For general unbounded statistics, the integrability condition in Definition~\ref{def:mean_identifiability} alone does not ensure differentiability of the mean. The following lemma reduces linear independence of signed measures to a finite matrix condition.

\begin{lemma}[Separation of signed measures by measurable sets]
\label{lem:finite_set_separation}
Let \(\zeta_1,\ldots,\zeta_m\) be finite signed measures on a measurable space \((\mathsf A,\mathcal A)\), with \(m\geq1\). They are linearly independent if and only if there exist measurable sets \(B_1,\ldots,B_m\in\mathcal A\) such that the matrix \([\zeta_j(B_i)]_{i,j=1}^m\) is nonsingular.
\end{lemma}

\begin{proof}
If the vectors \((\zeta_1(B),\ldots,\zeta_m(B))\), \(B\in\mathcal A\), do not span \(\R^m\), a nonzero vector orthogonal to their span gives a linear combination of the measures that vanishes on every measurable set. Otherwise, choose \(m\) such vectors forming a basis. The reverse implication follows by evaluating a vanishing linear combination on the selected measurable sets.
\end{proof}

\begin{theorem}[Local identifiability from recorded crossings]
\label{thm:local_identifiability}
Under Assumption~\ref{ass:signal_distribution_differentiability}, suppose \(r_{\mathcal H}\) is constant near \(\bm\theta_0\in\Theta\). The following statements are equivalent:
\begin{enumerate}
  \item \(\bm\theta_0\) is locally identifiable under \(\mathfrak S\).
  \item \(r_{\mathcal H}(\bm\theta_0)=M\).
  \item \(r_Y(\bm\theta_0)=M\), and \(\zeta\mapsto\zeta K_{\mathfrak S}\) is injective on the range of \(\left.D_{\bm\theta}P_{Y_{\bm\theta}}\right|_{\bm\theta=\bm\theta_0}\).
  \item There exist fixed measurable sets \(B_1,\ldots,B_M\in\B(\mathsf N_{\mathrm f}(\mathsf E))\) such that the probability map
  \begin{equation}
    \bm p(\bm\theta)\triangleq
    \bigl(\Upsilon(\bm\theta)(B_i)\bigr)_{i=1}^M
    \label{eq:recorded_probability_map}
  \end{equation}
  has a nonsingular Jacobian \(D_{\bm\theta}\bm p(\bm\theta_0)\).
\end{enumerate}
\end{theorem}

\begin{proof}
\emph{Equivalence of statements 2--4.} Statements 2 and 3 are equivalent by linearity. For each fixed measurable set \(B\), evaluation \(\zeta\mapsto\zeta(B)\) is bounded and linear, since \(|\zeta(B)|\leq\|\zeta\|_{\mathrm{TV}}\). Together with Proposition~\ref{prop:derivatives_through_sampling}, this gives continuous differentiability of \(\bm p\), with
\begin{equation}
  D_{\bm\theta}\bm p(\bm\theta)[\bm a]
  =\bigl((D_{\bm\theta}\Upsilon(\bm\theta)[\bm a])(B_i)\bigr)_{i=1}^M.
  \label{eq:probability_derivative_matrix}
\end{equation}
This holds for every \(\bm a\in\R^M\). Lemma~\ref{lem:finite_set_separation} applied to the parameter derivatives of \(\Upsilon\) gives the equivalence of statements 2 and 4.

\emph{Sufficiency.} Under statement 4, the inverse function theorem~\cite{lee2013smooth} gives a neighborhood on which \(\bm p\) is injective. Equality of distributions of recorded crossings implies equality of \(\bm p\), so statement 1 follows.

\emph{Necessity.} Suppose \(D_{\bm\theta}\Upsilon\) has constant rank \(r<M\) near \(\bm\theta_0\). If \(r=0\), then \(\Upsilon\) is constant on a sufficiently small ball, so \(\bm\theta_0\) is not locally identifiable.

For \(r>0\), Lemma~\ref{lem:finite_set_separation} gives \(r\) measurable sets such that the map \(\bm p^{(r)}:\Theta\to\R^r\) of their probabilities has derivative of rank \(r\) at \(\bm\theta_0\). A nonzero \(r\times r\) minor remains nonzero nearby by continuity. Since evaluation is linear and \(D_{\bm\theta}\Upsilon\) has rank \(r\), \(D_{\bm\theta}\bm p^{(r)}\) also has rank \(r\) throughout a sufficiently small neighborhood, with
\begin{equation*}
  \ker D_{\bm\theta}\Upsilon(\bm\theta)
  \subseteq\ker D_{\bm\theta}\bm p^{(r)}(\bm\theta).
\end{equation*}
Both kernels have dimension \(M-r\), so they are equal. After reordering the parameter coordinates, the first \(r\) columns of \(D_{\bm\theta}\bm p^{(r)}(\bm\theta_0)\) form a nonsingular matrix. The map
\begin{equation*}
  \bm\theta\mapsto
  \bigl(\bm p^{(r)}(\bm\theta),\theta_{r+1},\ldots,\theta_M\bigr)
\end{equation*}
therefore has a nonsingular Jacobian at \(\bm\theta_0\). The inverse function theorem supplies local coordinates \((\bm\alpha,\bm\beta)\in\R^r\times\R^{M-r}\) in which \(\bm p^{(r)}(\bm\theta)=\bm\alpha\). In a product of open balls in these coordinates, all derivatives of \(\Upsilon\) with respect to \(\bm\beta\) vanish. Thus \(\Upsilon\) depends only on \(\bm\alpha\), and its level set through \(\bm\theta_0\) is exactly the set with the same \(\bm\alpha\). This is a \(C^1\) manifold of dimension \(M-r>0\), contradicting local identifiability.
\end{proof}

Under Assumption~\ref{ass:signal_distribution_differentiability}, full rank is sufficient for local identifiability. The locally constant-rank condition is used for necessity: a rank deficiency then yields a positive-dimensional set of parameters with the same distribution, as shown in the proof.

The events in statement 4 are fixed as \(\bm\theta\) varies, but their choice may depend on the reference point \(\bm\theta_0\). A single collection valid throughout \(\Theta\) requires separate justification.

Statement 4 also gives local mean identifiability at \(\bm\theta_0\) with respect to \(\bm f(\nu)=(\ind_{B_i}(\nu))_{i=1}^M\), since \(\bm m_f=\bm p\) and the inverse function theorem applies. The efficiency of estimation based on these statistics remains a separate question.

\subsection{Conditions for Global Identifiability}
\label{subsec:global_identifiability}

Local identifiability can hold even when separated parameter values have the same distribution of recorded crossings. The next theorem characterizes global identifiability when the full-rank condition holds throughout \(\Theta\) and the image of \(\Upsilon\) satisfies the stated geometric assumptions. These image assumptions do not follow from full rank alone.

\begin{theorem}[Global identifiability from recorded crossings]
\label{thm:global_identifiability}
Under Assumption~\ref{ass:signal_distribution_differentiability}, suppose \(\Theta\) is connected and \(r_{\mathcal H}(\bm\theta)=M\) for every \(\bm\theta\in\Theta\).

Suppose also that \(\Upsilon(\Theta)\) is a simply connected \(M\)-dimensional \(C^1\) embedded submanifold without boundary of the space of finite signed measures on \(\mathsf N_{\mathrm f}(\mathsf E)\), equipped with the total variation norm. Compactness below is taken in the Euclidean topology on \(\Theta\) and the TV subspace topology on \(\Upsilon(\Theta)\). The following statements are equivalent:
\begin{enumerate}
  \item The parameter is globally identifiable on \(\Theta\) under \(\mathfrak S\).
  \item For every compact set \(\mathcal W\subseteq \Upsilon(\Theta)\),
  \begin{equation}
    \Upsilon^{-1}(\mathcal W)
    =\{\bm\theta\in\Theta:\Upsilon(\bm\theta)\in\mathcal W\}
    \label{eq:compact_recorded_distribution_preimages}
  \end{equation}
  is compact in \(\Theta\).
\end{enumerate}
Statement 2 means that \(\Upsilon:\Theta\to\Upsilon(\Theta)\) is proper.
\end{theorem}

\begin{proof}
\emph{Local invertibility.} By Proposition~\ref{prop:derivatives_through_sampling}, \(\Upsilon\) is continuously Fr\'echet differentiable. Its derivative takes values in the tangent space of \(\Upsilon(\Theta)\) at \(\Upsilon(\bm\theta)\). Since both this tangent space and \(\Theta\) have dimension \(M\), the rank condition makes the derivative an isomorphism onto that tangent space. The inverse function theorem in local coordinates therefore makes \(\Upsilon:\Theta\to \Upsilon(\Theta)\) a local \(C^1\) diffeomorphism.

\emph{Necessity of properness.} If statement 1 holds, \(\Upsilon\) is bijective onto \(\Upsilon(\Theta)\). Its local inverses give a continuous global inverse, so inverse images of compact sets are compact. Thus statement 2 follows.

\emph{Sufficiency of properness.} Conversely, under statement 2, \(\Upsilon\) is a proper local diffeomorphism between connected manifolds and hence a covering map~\cite{lee2013smooth}. Since \(\Upsilon(\Theta)\) is simply connected and \(\Theta\) is connected, this covering has one sheet; equivalently, \(\Upsilon\) is a global diffeomorphism. Hence \(\Upsilon\) is injective, which gives statement 1.
\end{proof}

\section{Minimum Number of Thresholds for Identifiability}
\label{sec:poisson_identifiability}

The minimum number of thresholds depends on the signal model and MT sampler. We apply the event-probability approach of Section~\ref{sec:identifiability} to a four-parameter Poisson signal model with a fixed known single-photon response and deterministic MT sampling. The construction uses four cumulative probabilities of the first recorded crossing time. We analyze their limits as the positive threshold tends to zero and transfer the resulting identifiability properties to sufficiently small positive thresholds. The local result applies near a fixed parameter value, while the global result provides a common threshold range over a bounded parameter domain. Both attain the lower bound \(N\geq1\) of the MT framework under their respective conditions.

\subsection{Signal Model and Sampler}
\label{sec:poisson_model_sampler}

\begin{definition}[Signal from nonhomogeneous Poisson arrivals]
\label{def:poisson_photon_signal}
Let \(\Theta\subseteq\R^4\) be nonempty and open, with \(\bm\theta=(\Lambda,t_0,\tau_{\mathrm{s,r}},\tau_{\mathrm{s,d}})\) satisfying \(\Lambda>0\) and \(0<\tau_{\mathrm{s,r}}<\tau_{\mathrm{s,d}}\) for every \(\bm\theta\in\Theta\). Fix the response parameters \(A>0\) and \(0<\tau_{\mathrm r}<\tau_{\mathrm d}\). For each \(\bm\theta\in\Theta\), let \(\{U_{j,\bm\theta}\}\) on \((\Omega,\mathcal F,\Prob)\) be the photon arrival times of a nonhomogeneous Poisson process on \(\R\) with photon-arrival intensity \(\lambda_{\bm\theta}\)~\cite{choong2009timing}. The photon-arrival intensity and fixed single-photon response are
\begin{equation}
  \begin{gathered}
    \lambda_{\bm\theta}(t)=\frac{\Lambda}{\tau_{\mathrm{s,d}}-\tau_{\mathrm{s,r}}}
      \left(e^{-(t-t_0)/\tau_{\mathrm{s,d}}}-e^{-(t-t_0)/\tau_{\mathrm{s,r}}}\right)
      \ind\{t\geq t_0\},\\
    h(t)=A\left(e^{-t/\tau_{\mathrm d}}-e^{-t/\tau_{\mathrm r}}\right)
      \ind\{t\geq0\}.
  \end{gathered}
  \label{eq:scintillation_example_model}
\end{equation}
Here, \(\Lambda=\int_{\R}\lambda_{\bm\theta}(u)\,du\) is the expected total photon count; \(t_0\), \(\tau_{\mathrm{s,r}}\), and \(\tau_{\mathrm{s,d}}\) are the onset time, rise time constant, and decay time constant of the photon-arrival intensity, respectively. For each \(\bm\theta\in\Theta\), define \(Y_{\bm\theta}\) almost surely by
\begin{equation}
  Y_{\bm\theta}(\omega)(t)\triangleq
  \sum_j h\bigl(t-U_{j,\bm\theta}(\omega)\bigr),
  \qquad t\in\mathcal T.
  \label{eq:poisson_photon_signal}
\end{equation}
\end{definition}

\begin{definition}[A deterministic MT sampler]
\label{def:deterministic_mt_sampler}
Fix a quantization step \(\Delta>0\). Consider a deterministic MT sampler \(\mathfrak S\) common to all parameter values, with threshold paths equal to the nominal thresholds (\(\widetilde{\mathbf V}=\mathbf V\) \(\Prob_{\mathcal C}\)-almost surely) and the time quantizer
\begin{equation*}
  r_\Delta(t)\triangleq\Delta\left\lfloor\frac{t}{\Delta}+\frac12\right\rfloor,
  \qquad t\in\R.
\end{equation*}
For \(\nu=\sum_{i=1}^{k_\nu}\delta_{(t_i,n_i,\sigma_i)}\in\mathsf N_{\mathrm f}(\mathsf E)\), where \(k_\nu=\nu(\mathsf E)\), define the deterministic timing and selection operators by
\begin{equation}
  \begin{gathered}
    \mathcal Q(\nu)\triangleq\sum_{i=1}^{k_\nu}
      \delta_{(r_\Delta(t_i),n_i,\sigma_i)},\\
    \mathcal H(\nu)\triangleq
      \sum_{\substack{1\leq i\leq k_\nu\\\sigma_i\in\{-1,+1\}}}
      \delta_{(t_i,n_i,\sigma_i)}.
  \end{gathered}
  \label{eq:deterministic_mt_sampler}
\end{equation}
For the fixed response in Definition~\ref{def:poisson_photon_signal}, set \(\Psi_0\equiv0\) on \(\R^0\), with value the zero signal path, and define, for \(k\geq1\) and \(\mathbf u\in\R^k\),
\begin{equation*}
  \begin{gathered}
    \Psi_k(\mathbf u)(t)\triangleq\sum_{j=1}^k h(t-u_j),
    \qquad t\in\mathcal T,\\
    \mathsf G_h\triangleq\{0\}\cup\bigcup_{k\geq1}\Psi_k(\R^k)
    \subseteq\mathsf X.
  \end{gathered}
\end{equation*}
Set \(\mathsf D_{\mathcal C}=\mathsf G_h\times\{\mathbf V\}\) and \(\mathsf X_{\mathcal C}=\mathsf G_h\). The following proposition verifies the admissible domain and the common admissibility condition
\begin{equation}
  P_{Y_{\bm\theta}}(\mathsf X_{\mathcal C})=1,
  \qquad \bm\theta\in\Theta.
  \label{eq:poisson_common_admissibility}
\end{equation}
The stochastic MT sampling operator in \eqref{eq:stochastic_mt_sampling_operator} takes the form
\begin{equation*}
  \mathcal S(\cdot,\omega_{\mathrm s})
  =\mathcal H\circ\mathcal Q\circ\mathcal C(\cdot,\mathbf V)
\end{equation*}
on \(\mathsf X_{\mathcal C}\), for \(\Prob_{\mathrm s}\)-almost every \(\omega_{\mathrm s}\).
\end{definition}

The four unknown parameters govern the photon-arrival intensity; the single-photon response \(h\) and the sampler are fixed and known. Thus the maps \(\Psi_k\) and the sampling kernel are independent of \(\bm\theta\). The next proposition verifies the common admissible domain and the total variation differentiability required by the identifiability theorems.

\begin{proposition}[Admissibility and differentiability under Poisson arrivals]
\label{prop:poisson_distribution_differentiability}
For the signal and sampler in Definitions~\ref{def:poisson_photon_signal} and~\ref{def:deterministic_mt_sampler}, \(\mathsf D_{\mathcal C}\) is an admissible domain, and \(\mathsf X_{\mathcal C}=\mathsf G_h\) satisfies \eqref{eq:poisson_common_admissibility} independently of the intensity parameters. Moreover, \(Y_{\bm\theta}\in\mathfrak Y\) for every \(\bm\theta\in\Theta\), and Assumption~\ref{ass:signal_distribution_differentiability} holds with \(M=4\). Consequently, \(\bm\theta\mapsto P_{\mathcal P_{\bm\theta}}\) is continuously Fr\'echet differentiable in the total variation norm.
\end{proposition}

\begin{proof}
\emph{Signal integrability and common admissibility.} The response \(h\) is continuous with \(0\leq h\leq A\), and the expected total photon count is \(\Lambda\). Thus \(Y_{\bm\theta}\) is a measurable \(\mathsf X\)-valued random element, and
\begin{equation*}
  \Exp\|Y_{\bm\theta}\|_\infty
  \leq\int_{\R}\lambda_{\bm\theta}(u)
  \sup_{t\in\mathcal T}|h(t-u)|\,du\leq A\Lambda<\infty.
\end{equation*}
Thus \(Y_{\bm\theta}\in\mathfrak Y\).

For each \(k\geq1\), the finite-sum map \(\Psi_k:\R^k\to\mathsf X\) is continuous in the uniform norm. Since \(\R^k\) is a countable union of compact sets, \(\mathsf G_h\) is a countable union of compact sets in the Skorokhod topology and is Borel. Between successive arrival times, each path minus a constant threshold has an analytic extension across the interval endpoints. It is either identically zero or has finitely many zeros on the interval closure. This gives the required finite comparison-state partition and verifies the admissible domain in Definition~\ref{def:deterministic_mt_sampler}. The finite total photon count gives \eqref{eq:poisson_common_admissibility} for every parameter value satisfying Definition~\ref{def:poisson_photon_signal}.

\emph{\(L^1\) differentiability of the intensity.} For fixed intensity rise and decay time constants, write \(\lambda_{\bm\theta}(t)=\Lambda g(t-t_0)\), where
\begin{equation*}
  g(s)=\frac{e^{-s/\tau_{\mathrm{s,d}}}-e^{-s/\tau_{\mathrm{s,r}}}}
    {\tau_{\mathrm{s,d}}-\tau_{\mathrm{s,r}}}\ind_{\{s\geq0\}}.
\end{equation*}
Since \(g(0)=0\), its zero extension to negative times is absolutely continuous, with \(g,g'\in L^1(\R)\). For \(\varepsilon\ne0\),
\begin{equation*}
  \begin{aligned}
    &\left\|\frac{g(\cdot-\varepsilon)-g}{\varepsilon}+g'\right\|_{L^1}\\
    &\qquad\leq\int_0^1
      \|g'(\cdot-r\varepsilon)-g'\|_{L^1}\,dr
      \longrightarrow0
  \end{aligned}
\end{equation*}
as \(\varepsilon\to0\), by continuity of translations in \(L^1\). Hence the onset derivative is represented by the \(L^1\) function \(D_{t_0}\lambda_{\bm\theta}(t)=-\Lambda g'(t-t_0)\) almost everywhere.

On a compact parameter neighborhood \(\mathcal K\subset\Theta\), set \(a=\min_{\bm\theta\in\mathcal K}t_0\) and \(b=\max_{\bm\theta\in\mathcal K}\tau_{\mathrm{s,d}}\). The time constants and their difference are bounded away from zero, and all parameters are bounded. The explicit derivatives therefore admit representatives such that, for some \(C_{\mathcal K}>0\),
\begin{equation*}
  \begin{gathered}
    \rho_{\mathcal K}(u)\triangleq
      C_{\mathcal K}(1+u-a)e^{-(u-a)/b}\ind\{u\geq a\},\\
    |\lambda_{\bm\theta}(u)|+
      \|D_{\bm\theta}\lambda_{\bm\theta}(u)\|_2
      \leq\rho_{\mathcal K}(u),
      \qquad \bm\theta\in\mathcal K,
  \end{gathered}
\end{equation*}
almost everywhere, with \(\rho_{\mathcal K}\in L^1(\R)\). At each fixed parameter value, these derivatives vary pointwise continuously for \(u\ne t_0\). Dominated convergence gives their \(L^1\) continuity. Thus \(\bm\theta\mapsto\lambda_{\bm\theta}\) is continuously Fr\'echet differentiable in \(L^1(\R)\).

\emph{Total variation differentiability of the signal distribution.} For \(k\geq0\) and \(\mathbf u\in\R^k\), define
\begin{equation}
  w_{\bm\theta,k}(\mathbf u)
  \triangleq\frac{e^{-\Lambda}}{k!}
    \prod_{j=1}^k\lambda_{\bm\theta}(u_j),
  \label{eq:poisson_configuration_density}
\end{equation}
with the empty product equal to one, so \(w_{\bm\theta,0}=e^{-\Lambda}\). For \(B\in\B(\mathsf X)\), the Poisson expansion~\cite{last2018lectures} gives
\begin{equation*}
  P_{Y_{\bm\theta}}(B)
  =\sum_{k=0}^{\infty}
  \int_{\R^k}
  \ind_B\!\left(\Psi_k(\mathbf u)\right)
  w_{\bm\theta,k}(\mathbf u)\,d\mathbf u,
\end{equation*}
where \(d\mathbf u=du_1\cdots du_k\) for \(k\geq1\), and integration over \(\R^0\) evaluates the integrand at the empty tuple. The \(k=0\) term is therefore \(e^{-\Lambda}\ind_B(0)\).

For a direction \(\bm a\in\R^4\), let \(a_\Lambda\) denote its component corresponding to \(\Lambda\). For \(k\geq1\), differentiation in \(L^1(\R^k)\) gives
\begin{equation}
  \begin{aligned}
    &D_{\bm\theta}w_{\bm\theta,k}(\mathbf u)[\bm a]
      =\frac{e^{-\Lambda}}{k!}\Biggl[
      -a_\Lambda\prod_{j=1}^k\lambda_{\bm\theta}(u_j)\\
    &\qquad+\sum_{m=1}^k D_{\bm\theta}\lambda_{\bm\theta}(u_m)[\bm a]
      \prod_{\substack{1\leq j\leq k\\j\ne m}}\lambda_{\bm\theta}(u_j)\Biggr].
  \end{aligned}
  \label{eq:poisson_configuration_density_derivative}
\end{equation}
For \(k=0\), \(D_{\bm\theta}w_{\bm\theta,0}[\bm a]=-a_\Lambda e^{-\Lambda}\).

On a sufficiently small parameter neighborhood, choose \(L\geq1\) and \(C>0\) such that \(\|\lambda_{\bm\theta}\|_{L^1}\leq L\) and \(\|D_{\bm\theta}\lambda_{\bm\theta}[\bm a]\|_{L^1}\leq C\) for every unit direction \(\bm a\). For \(k\geq1\), the density and its derivative then satisfy
\begin{equation*}
  \begin{aligned}
    \|w_{\bm\theta,k}\|_{L^1}&\leq\frac{L^k}{k!},\\
    \|D_{\bm\theta}w_{\bm\theta,k}[\bm a]\|_{L^1}
      &\leq\frac{L^k+kCL^{k-1}}{k!}.
  \end{aligned}
\end{equation*}
Each summand of the signal distribution is the image of \(w_{\bm\theta,k}(\mathbf u)\,d\mathbf u\) under the fixed measurable map \(\Psi_k\). These images inherit continuous differentiability and the same bounds in TV. Both bounds are summable over \(k\), and the zero-count term is continuously differentiable, so the distribution series can be differentiated termwise in TV. Restriction to the common \(\mathsf X_{\mathcal C}\) preserves this differentiability, and Proposition~\ref{prop:derivatives_through_sampling} gives the result for the distribution of recorded crossings.
\end{proof}

For \(N=1\), write \(\mathfrak S_V\) for the sampler in Definition~\ref{def:deterministic_mt_sampler} with threshold \(V_1=V>0\), and write \(\mathcal P_{\bm\theta}^{(V)}\) for its recorded crossings. Define the map from parameters to the distribution of recorded crossings by
\begin{equation}
  \Upsilon_V(\bm\theta)
  \triangleq P_{\mathcal P_{\bm\theta}^{(V)}}
  =P_{Y_{\bm\theta}}K_{\mathfrak S_V}.
  \label{eq:threshold_recorded_distribution_map}
\end{equation}
Parameter derivatives are taken at fixed \(V\); varying \(V\) changes the sampler.

Following the first recorded crossing-time definition in~\cite{qiu2026sampling}, fix the bounded open recorded-time interval \(I=(T_--\Delta,T_++\Delta)\), which contains every quantized time from \((T_-,T_+)\). For \(\nu\in\mathsf N_{\mathrm f}(\mathsf E)\), set
\begin{equation*}
  \tau_{1,I}(\nu)\triangleq
  \inf\{t\in I:\nu(\{t\}\times\{1\}\times\{-1,+1\})>0\},
\end{equation*}
with \(\inf\varnothing=\infty\). For \(t\in\R\), let \(D_t=(I\cap({-\infty},t])\times\{1\}\times\{-1,+1\}\). Then
\begin{equation*}
  \{\nu:\tau_{1,I}(\nu)\leq t\}
  =\{\nu:\nu(D_t)>0\}
\end{equation*}
is a measurable count event, so \(\tau_{1,I}\) is measurable. Write \(\tau_{\bm\theta}^{(V)}=\tau_{1,I}(\mathcal P_{\bm\theta}^{(V)})\) and \(F_{V,\bm\theta}(t)=\Prob\{\tau_{\bm\theta}^{(V)}\leq t\}\) for the first recorded crossing time and its unconditional cumulative distribution function (CDF). Realizations with no recorded crossings have \(\tau_{\bm\theta}^{(V)}=\infty\), and
\begin{equation*}
  \lim_{t\to+\infty}F_{V,\bm\theta}(t)
  =\Prob\{\tau_{\bm\theta}^{(V)}<\infty\}.
\end{equation*}
The remaining probability mass is at \(\infty\).

\subsection{Local Identifiability}
\label{sec:poisson_local_identifiability}

Four cumulative probabilities of the first recorded crossing time suffice for local identifiability. The condition below requires the fixed quantization step and observation window to admit four distinct analog-time cutoffs after the reference onset. The threshold bound \(V_*\) may depend on \(\bm\theta_0\) and these cutoffs.

\begin{corollary}[Minimum number of thresholds for local identifiability]
\label{cor:minimum_thresholds_local_identifiability}
Consider the signal in Definition~\ref{def:poisson_photon_signal} and the deterministic MT sampler in Definition~\ref{def:deterministic_mt_sampler}. Fix \(\bm\theta_0\in\Theta\), and denote its onset time by \(t_{0,0}\). Suppose there are integers \(\kappa_1<\kappa_2<\kappa_3<\kappa_4\) such that
\begin{equation}
  \begin{gathered}
    c_i\triangleq\min\{\Delta\kappa_i+\Delta/2,T_+\},\qquad i=1,\ldots,4,\\
    T_-<t_{0,0}<c_1<c_2<c_3<c_4\leq T_+.
  \end{gathered}
  \label{eq:four_recording_cutoffs}
\end{equation}
There exists \(V_*>0\) such that every sampler with \(N=1\) and threshold \(V_1=V\in(0,V_*)\) makes \(\bm\theta_0\) locally identifiable. Thus the minimum number of thresholds is one.
\end{corollary}

\begin{proof}
\emph{Recorded events and analog-time cutoffs.} For \(\mathfrak S_V\), let
\begin{equation*}
  \begin{aligned}
    B_i&\triangleq\{\nu:\nu(({-\infty},\Delta\kappa_i]\times\{1\}\times\{+1\})>0\},\\
    p_i(V,\bm\theta)&\triangleq P_{\mathcal P_{\bm\theta}^{(V)}}(B_i),
    \qquad i=1,\ldots,4,\\
    \bm p_V(\bm\theta)&\triangleq\bigl(p_i(V,\bm\theta)\bigr)_{i=1}^4.
  \end{aligned}
\end{equation*}
Here \(\nu\in\mathsf N_{\mathrm f}(\mathsf E)\), and \(B_i\in\B(\mathsf N_{\mathrm f}(\mathsf E))\). The quantity \(p_i(V,\bm\theta)\) is the probability of at least one recorded crossing with direction mark \(+1\) at or before the recorded-time cutoff \(\Delta\kappa_i\). Since \(\kappa_i\in\mathbb Z\),
\begin{equation*}
  r_\Delta(t)\leq\Delta\kappa_i\quad\Longleftrightarrow\quad t<\Delta\kappa_i+\Delta/2.
\end{equation*}
Analog crossing times lie in \((T_-,T_+)\), so \(B_i\) occurs exactly when an analog crossing with direction mark \(+1\) occurs before the analog-time cutoff \(c_i\). Thus \(\Delta\kappa_i\) and \(c_i\) are cutoffs in recorded and analog time, respectively. Restrict \(\bm\theta\) to a neighborhood of \(\bm\theta_0\) on which \(T_-<t_0<c_1\). Since the signal starts below \(V>0\) and the quantizer preserves time order, the first recorded crossing time \(\tau_{\bm\theta}^{(V)}\), when finite, is the recorded time of an upward crossing. Consequently, \(p_i(V,\bm\theta)=F_{V,\bm\theta}(\Delta\kappa_i)\).

\emph{Limit as \(V\downarrow0\).} The response is zero before arrival and strictly increasing immediately afterwards. For \(k\geq1\), restrict attention to arrival configurations \(\mathbf u\in(T_-,\infty)^k\); the Poisson densities and their parameter derivatives vanish almost everywhere outside this set on the chosen parameter neighborhood. For almost every such configuration with an arrival before \(c_i\), a short interval after the first arrival contains no further arrivals, and the signal rises strictly from zero on that interval. Thus an analog crossing with direction mark \(+1\) occurs before \(c_i\) for every sufficiently small positive \(V\). Consequently, as \(V\downarrow0\),
\begin{equation}
  p_i(V,\bm\theta)\longrightarrow
  p_i^{\mathrm{lim}}(\bm\theta)\triangleq
  1-\exp\!\left(-\int_{t_0}^{c_i}\lambda_{\bm\theta}(u)\,du\right).
  \label{eq:low_threshold_probability_limit}
\end{equation}
Write \(\bm p_{\mathrm{lim}}(\bm\theta)\triangleq(p_i^{\mathrm{lim}}(\bm\theta))_{i=1}^4\). This limit as \(V\downarrow0\) differs from sampling at \(V=0\): the first transition from equality to a positive signal then has direction mark \(+\tfrac12\) and is removed by \(\mathcal H\). Configurations with no arrival before \(c_i\) contribute zero; arrival coincidences and arrivals at the cutoffs have Lebesgue measure zero.

\emph{Convergence of parameter derivatives.} To transfer Jacobian nonsingularity from the limiting map to \(\bm p_V\), we also need convergence of the parameter derivatives. For \(k\geq1\), let \(I_{i,V,k}(\mathbf u)\) be the indicator of \(B_i\) for the recorded crossings generated by \(\Psi_k(\mathbf u)\) under \(\mathfrak S_V\), and set \(I_{i,V,0}=0\). These indicators are independent of \(\bm\theta\), so Proposition~\ref{prop:poisson_distribution_differentiability} gives
\begin{equation*}
  D_{\bm\theta}p_i(V,\bm\theta_0)
  =\sum_{k=1}^{\infty}\int_{\R^k}
    I_{i,V,k}(\mathbf u)
    D_{\bm\theta}w_{\bm\theta_0,k}(\mathbf u)\,d\mathbf u.
\end{equation*}
On the supports of the density derivatives, \(I_{i,V,k}\) converges almost everywhere to \(\ind\{\min_j u_j<c_i\}\). Dominated convergence, first in each integral and then in the series using the summable bounds of Proposition~\ref{prop:poisson_distribution_differentiability}, therefore gives
\begin{equation}
  D_{\bm\theta}\bm p_V(\bm\theta_0)
  \longrightarrow
  D_{\bm\theta}\bm p_{\mathrm{lim}}(\bm\theta_0).
  \label{eq:low_threshold_derivative_limit}
\end{equation}

\emph{Jacobian nonsingularity.} We now show that the limiting Jacobian is nonsingular. The normalized photon-arrival intensity defines the photon-arrival CDF
\begin{equation*}
  \begin{gathered}
    \Gamma(s;\tau_{\mathrm{s,r}},\tau_{\mathrm{s,d}})\triangleq
    \frac1\Lambda\int_{t_0}^{t_0+s}\lambda_{\bm\theta}(u)\,du\\
    =1-\frac{\tau_{\mathrm{s,d}}e^{-s/\tau_{\mathrm{s,d}}}
      -\tau_{\mathrm{s,r}}e^{-s/\tau_{\mathrm{s,r}}}}
      {\tau_{\mathrm{s,d}}-\tau_{\mathrm{s,r}}},\qquad s\geq0.
  \end{gathered}
\end{equation*}
Here \(s\) is time relative to onset, and \(p_i^{\mathrm{lim}}=1-e^{-\Lambda\Gamma(c_i-t_0;\tau_{\mathrm{s,r}},\tau_{\mathrm{s,d}})}\). Removing nonzero row and column factors from \(D_{\bm\theta}\bm p_{\mathrm{lim}}\) leaves the evaluation matrix of
\begin{equation*}
  \Gamma,\quad -D_s\Gamma,\quad
  D_{\tau_{\mathrm{s,r}}}\Gamma,\quad D_{\tau_{\mathrm{s,d}}}\Gamma
\end{equation*}
at the four distinct positive points \(s_i=c_i-t_0\). To verify linear independence, suppose a linear combination of these functions vanishes identically. Its constant term forces the coefficient of \(\Gamma\) to vanish. The terms \(se^{-s/\tau_{\mathrm{s,r}}}\) and \(se^{-s/\tau_{\mathrm{s,d}}}\) then force the coefficients of the respective time-constant derivatives to vanish, leaving only a multiple of the nonzero function \(-D_s\Gamma\). Thus all coefficients vanish. To establish nonsingularity of the evaluation matrix, we combine this independence with a bound on the number of zeros. All four functions vanish at zero and belong to
\begin{equation*}
  \operatorname{span}\{1,e^{-s/\tau_{\mathrm{s,r}}},se^{-s/\tau_{\mathrm{s,r}}},
    e^{-s/\tau_{\mathrm{s,d}}},se^{-s/\tau_{\mathrm{s,d}}}\}.
\end{equation*}
A nonzero function in this space has at most four distinct zeros. To see this, suppose \(\varphi\) has five distinct zeros. Then \(\varphi'\), and hence \(\psi(s)=e^{s/\tau_{\mathrm{s,r}}}\varphi'(s)\), has at least four. The function \(\psi''\) is a polynomial of degree at most one times \(e^{(1/\tau_{\mathrm{s,r}}-1/\tau_{\mathrm{s,d}})s}\). Rolle's theorem gives at least two distinct zeros of \(\psi''\), forcing \(\psi''\equiv0\). Thus \(\psi\) is affine, and its four zeros imply \(\psi\equiv0\). It follows that \(\varphi\) is constant, and its zeros give \(\varphi\equiv0\). A singular evaluation matrix would thus give a nonzero combination with the five zeros \(0,s_1,\ldots,s_4\), a contradiction.

Thus \(D_{\bm\theta}\bm p_{\mathrm{lim}}(\bm\theta_0)\) is nonsingular, and \eqref{eq:low_threshold_derivative_limit} makes \(D_{\bm\theta}\bm p_V(\bm\theta_0)\) nonsingular for every sufficiently small positive \(V\). Hence \(r_{\mathcal H}(\bm\theta_0)=4\) for \(\mathfrak S_V\), and continuity makes this rank constant near \(\bm\theta_0\). Theorem~\ref{thm:local_identifiability} gives local identifiability.
\end{proof}

With \(B_i\) as in the proof, \(\bm f(\nu)=(\ind_{B_i}(\nu))_{i=1}^4\) has mean \(\bm p_V(\bm\theta)\) under \(\mathfrak S_V\); thus Corollary~\ref{cor:minimum_thresholds_local_identifiability} also establishes local mean identifiability at \(\bm\theta_0\) with respect to this fixed statistic for every sufficiently small positive \(V\).

\subsection{Global Identifiability}
\label{sec:poisson_global_identifiability}

The global result provides one threshold range valid throughout a bounded convex parameter domain. The same four cutoffs must apply to every parameter value, with the first cutoff later than every onset in the closure of the domain. The proof establishes injectivity of \(\bm p_V\) directly and then connects the result to Theorem~\ref{thm:global_identifiability}.

\begin{corollary}[Minimum number of thresholds for global identifiability]
\label{cor:minimum_thresholds_global_identifiability}
Consider the signal in Definition~\ref{def:poisson_photon_signal} and the deterministic MT sampler in Definition~\ref{def:deterministic_mt_sampler}. Suppose \(\Theta\) is nonempty, bounded, open, and convex. Fix integers \(\kappa_1<\kappa_2<\kappa_3<\kappa_4\), and set \(c_i=\min\{\Delta\kappa_i+\Delta/2,T_+\}\). Assume \(c_1<c_2<c_3<c_4\leq T_+\) and
\begin{equation}
  \overline{\Theta}\subset
  \left\{\bm\theta\in\R^4:
  \begin{gathered}
    \Lambda>0,\quad T_-<t_0<c_1,\\
    0<\tau_{\mathrm{s,r}}<\tau_{\mathrm{s,d}}
  \end{gathered}
  \right\}.
  \label{eq:global_poisson_parameter_bounds}
\end{equation}
There exists \(V_*>0\) such that every sampler with \(N=1\) and threshold \(V_1=V\in(0,V_*)\) makes the parameter globally identifiable on \(\Theta\). Thus the minimum number of thresholds is one.
\end{corollary}

\begin{proof}
Use \(B_i\), \(\bm p_V\), and \(\bm p_{\mathrm{lim}}\) from the local proof.
By \eqref{eq:global_poisson_parameter_bounds}, the inequalities \(\Lambda>0\), \(T_-<t_0<c_1\), and \(0<\tau_{\mathrm{s,r}}<\tau_{\mathrm{s,d}}\) hold on an open neighborhood of \(\overline{\Theta}\). The common set of admissible signal paths \(\mathsf G_h\) allows the same signal and sampling formulas to define these maps on that neighborhood, where Proposition~\ref{prop:poisson_distribution_differentiability} applies.

\emph{Injectivity of the limiting probability map.} We show that \(\bm p_{\mathrm{lim}}\) is injective on \(\overline{\Theta}\). For \(t\geq t_0\), define the cumulative photon-arrival intensity by
\begin{equation*}
  \begin{aligned}
    \Lambda_{\bm\theta}(t)
      &\triangleq\int_{t_0}^{t}\lambda_{\bm\theta}(u)\,du\\
      &=\Lambda\Gamma(t-t_0;\tau_{\mathrm{s,r}},\tau_{\mathrm{s,d}}).
  \end{aligned}
\end{equation*}
Thus \(\Lambda_{\bm\theta}(t)\) is the expected number of arrivals by \(t\), while \(\Lambda=\int_{\R}\lambda_{\bm\theta}(u)\,du\) is the expected total photon count.
If \(\bm p_{\mathrm{lim}}(\bm\theta)=\bm p_{\mathrm{lim}}(\bm\theta')\), then \(\chi(t)\triangleq\Lambda_{\bm\theta}(t)-\Lambda_{\bm\theta'}(t)\) vanishes at all four \(c_i\). After both onset times, \(\chi'\) is a linear combination of at most four distinct real exponential functions. Unless identically zero, it has at most three distinct zeros: this follows by induction using Rolle's theorem after division by one exponential.

If the onset times agree, \(\chi\) also vanishes at their common onset, so Rolle's theorem gives at least four zeros of \(\chi'\). If \(t_0<t_0'\), then \(\chi(t_0')>0\) and \(\chi'(t_0')>0\). Since \(\chi(c_1)=0\), the mean value theorem and continuity give a zero of \(\chi'\) in \((t_0',c_1)\). The three intervals \((c_i,c_{i+1})\) give three more zeros, a contradiction. Interchanging the parameters handles \(t_0>t_0'\). Hence the onsets agree and the intensities coincide. Linear independence of exponentials, together with \(\tau_{\mathrm{s,r}}<\tau_{\mathrm{s,d}}\), then gives equality of both time constants and \(\Lambda\). Thus \(\bm\theta=\bm\theta'\).

\emph{Uniform convergence of probabilities and derivatives.} We strengthen the limits in Corollary~\ref{cor:minimum_thresholds_local_identifiability} to
\begin{equation}
  \sup_{\bm\theta\in\overline{\Theta}}
  \left(
    \|\bm p_V-\bm p_{\mathrm{lim}}\|_2+
    \|D_{\bm\theta}\bm p_V-D_{\bm\theta}\bm p_{\mathrm{lim}}\|_2
  \right)\longrightarrow0
  \quad(V\downarrow0),
  \label{eq:uniform_low_threshold_limit}
\end{equation}
where all maps inside the supremum are evaluated at \(\bm\theta\). Using the extension to a neighborhood of \(\overline{\Theta}\), apply the envelope construction in Proposition~\ref{prop:poisson_distribution_differentiability} with \(\mathcal K=\overline{\Theta}\) and write \(\rho=\rho_{\mathcal K}\). Since \(\min_{\bm\theta\in\overline{\Theta}}t_0>T_-\), this nonnegative integrable envelope vanishes on \(({-\infty},T_-]\) and satisfies
\begin{equation*}
  \sup_{\bm\theta\in\overline{\Theta}}
  \left(|\lambda_{\bm\theta}(u)|+
  \|D_{\bm\theta}\lambda_{\bm\theta}(u)\|_2\right)
  \leq\rho(u)
  \quad\text{a.e.}
\end{equation*}
By \eqref{eq:poisson_configuration_density_derivative}, for \(k\geq1\), the density \(w_{\bm\theta,k}\) and the Euclidean norm of its parameter derivative are each bounded, uniformly over \(\overline\Theta\), by \((k+1)\prod_{j=1}^k\rho(u_j)/k!\).

Retain \(I_{i,V,k}\) from the local proof and, for \(k\geq1\), define
\begin{equation*}
  I_{i,k}^{\mathrm{lim}}(\mathbf u)
  \triangleq\ind\!\left\{\min_{1\leq j\leq k}u_j<c_i\right\},
  \qquad \mathbf u\in\R^k.
\end{equation*}
Set \(I_{i,0}^{\mathrm{lim}}=0\). The superscript \(\mathrm{lim}\) denotes the limiting cutoff condition as \(V\downarrow0\). Both indicators are common to all parameter values. Wherever \(\prod_j\rho(u_j)>0\), all arrival times exceed \(T_-\). The configuration argument in Corollary~\ref{cor:minimum_thresholds_local_identifiability} therefore gives \(I_{i,V,k}\to I_{i,k}^{\mathrm{lim}}\) almost everywhere on this set as \(V\downarrow0\).

The probabilities have the representations
\begin{equation*}
  \begin{aligned}
    p_i(V,\bm\theta)
      &=\sum_{k=1}^{\infty}\int_{\R^k}
        I_{i,V,k}(\mathbf u)w_{\bm\theta,k}(\mathbf u)\,d\mathbf u,\\
    p_i^{\mathrm{lim}}(\bm\theta)
      &=\sum_{k=1}^{\infty}\int_{\R^k}
        I_{i,k}^{\mathrm{lim}}(\mathbf u)w_{\bm\theta,k}(\mathbf u)\,d\mathbf u.
  \end{aligned}
\end{equation*}
Their parameter derivatives are obtained by replacing \(w_{\bm\theta,k}\) with \(D_{\bm\theta}w_{\bm\theta,k}\). Consequently, for each \(i=1,\ldots,4\),
\begin{equation}
  \begin{aligned}
    &\sup_{\bm\theta\in\overline\Theta}\Bigl(
      |p_i-p_i^{\mathrm{lim}}|
      +\|D_{\bm\theta}p_i-D_{\bm\theta}p_i^{\mathrm{lim}}\|_2\Bigr)\\
    &\leq2\sum_{k=1}^{\infty}\frac{k+1}{k!}
      \int_{\R^k}|I_{i,V,k}-I_{i,k}^{\mathrm{lim}}|
      \prod_{j=1}^k\rho(u_j)\,d\mathbf u\longrightarrow0,
  \end{aligned}
  \label{eq:uniform_recording_probability_derivative_bound}
\end{equation}
as \(V\downarrow0\), where \(p_i\) and \(p_i^{\mathrm{lim}}\) are evaluated at \((V,\bm\theta)\) and \(\bm\theta\), respectively, and both indicators at \(\mathbf u\). The bound is uniform in \(\bm\theta\). Dominated convergence applies to each integral and then to the series, since the indicator differences are bounded by one and
\begin{equation*}
  \sum_{k=1}^{\infty}\frac{k+1}{k!}\|\rho\|_{L^1}^k<\infty.
\end{equation*}
Combining the four component bounds gives \eqref{eq:uniform_low_threshold_limit}, including convergence in the induced matrix norm for the derivatives.

\emph{Persistence of injectivity at positive thresholds.} We treat nearby and separated parameter pairs using uniform derivative and probability convergence, respectively. Put \(J_0(\bm\theta)=D_{\bm\theta}\bm p_{\mathrm{lim}}(\bm\theta)\). The proof of Corollary~\ref{cor:minimum_thresholds_local_identifiability} shows that \(J_0\) is nonsingular throughout \(\overline{\Theta}\), so
\begin{equation*}
  \gamma_*\triangleq
    \min_{\bm\theta\in\overline{\Theta}}\sigma_{\min}(J_0(\bm\theta))>0.
\end{equation*}
Uniform continuity gives \(\delta>0\) such that
\(\|J_0(\bm\theta')-J_0(\bm\theta)\|_2<\gamma_*/4\)
whenever \(\bm\theta,\bm\theta'\in\overline{\Theta}\) and
\(\|\bm\theta'-\bm\theta\|_2<\delta\). For all sufficiently small \(V>0\), \eqref{eq:uniform_low_threshold_limit} also gives
\begin{equation*}
  \sup_{\bm\theta\in\overline{\Theta}}
    \|D_{\bm\theta}\bm p_V(\bm\theta)-J_0(\bm\theta)\|_2
    <\gamma_*/4.
\end{equation*}
For \(\bm\theta,\bm\theta'\in\overline{\Theta}\) with
\(\|\bm\theta'-\bm\theta\|_2<\delta\), set
\(\bm a=\bm\theta'-\bm\theta\). Convexity keeps the joining segment in \(\overline{\Theta}\), and the two bounds imply
\(\|D_{\bm\theta}\bm p_V(\bm\theta+s\bm a)-J_0(\bm\theta)\|_2<\gamma_*/2\)
for \(0\leq s\leq1\). Integration along that segment gives
\begin{equation*}
  \begin{aligned}
    &\|\bm p_V(\bm\theta')-\bm p_V(\bm\theta)-J_0(\bm\theta)\bm a\|_2\\
    &\quad\leq\int_0^1
      \|D_{\bm\theta}\bm p_V(\bm\theta+s\bm a)-J_0(\bm\theta)\|_2
      \|\bm a\|_2\,ds\\
    &\quad\leq\frac{\gamma_*}{2}\|\bm a\|_2.
  \end{aligned}
\end{equation*}
Hence
\begin{equation*}
  \|\bm p_V(\bm\theta')-\bm p_V(\bm\theta)\|_2
  \geq\frac{\gamma_*}{2}\|\bm\theta'-\bm\theta\|_2
  \quad\text{if }\|\bm\theta'-\bm\theta\|_2<\delta.
\end{equation*}

If pairs in \(\overline{\Theta}\) at distance at least \(\delta\) exist, compactness and injectivity of \(\bm p_{\mathrm{lim}}\) give
\begin{equation*}
  m_\delta\triangleq
    \min_{\substack{\bm\theta,\bm\theta'\in\overline{\Theta}\\
      \|\bm\theta'-\bm\theta\|_2\geq\delta}}
    \|\bm p_{\mathrm{lim}}(\bm\theta')-\bm p_{\mathrm{lim}}(\bm\theta)\|_2>0.
\end{equation*}
By \eqref{eq:uniform_low_threshold_limit}, further restricting \(V>0\) ensures
\(\sup_{\overline{\Theta}}\|\bm p_V-\bm p_{\mathrm{lim}}\|_2<m_\delta/3\).
The triangle inequality then gives
\(\|\bm p_V(\bm\theta')-\bm p_V(\bm\theta)\|_2>m_\delta/3\)
for these pairs. Choose \(V_*>0\) so that the derivative bound and, when needed, the probability bound hold for \(0<V<V_*\). Then \(\bm p_V\) is injective on \(\overline{\Theta}\) throughout this threshold range. The derivative bound also gives
\(\sigma_{\min}(D_{\bm\theta}\bm p_V(\bm\theta))\geq3\gamma_*/4>0\)
throughout \(\overline{\Theta}\).

Fix such a \(V\). Equality of distributions of recorded crossings implies equality of \(\bm p_V\), and hence of the parameters, proving global identifiability.

\emph{Connection to the general global criterion.} We now verify the rank, image, and properness conditions of Theorem~\ref{thm:global_identifiability}. On \(\Theta\), the map \(\bm p_V\) is a \(C^1\) diffeomorphism onto its open image. Define the bounded linear evaluation map \(E\mu=(\mu(B_i))_{i=1}^4\) on finite signed measures. Since \(E\Upsilon_V=\bm p_V\) and \(D_{\bm\theta}\bm p_V\) is nonsingular, \(r_{\mathcal H}=4\) for \(\mathfrak S_V\), and \(E\) has a bounded linear right inverse \(\mathcal L\). The decomposition into \(\operatorname{ran}\mathcal L\) and \(\ker E\) gives
\begin{equation*}
  \begin{aligned}
    \Upsilon_V(\bm p_V^{-1}(\bm y))
    &=\mathcal L\bm y+(\Id-\mathcal L E)\Upsilon_V(\bm p_V^{-1}(\bm y)),\\
    &\hspace{2em}\bm y\in\bm p_V(\Theta).
  \end{aligned}
\end{equation*}
The second term takes values in \(\ker E\) and is \(C^1\), so \(\Upsilon_V(\Theta)\) is a \(C^1\) graph over \(\bm p_V(\Theta)\). It is therefore a four-dimensional embedded submanifold without boundary, homeomorphic to the convex set \(\Theta\), and hence simply connected.

Finally, the inverse map is
\begin{equation*}
  \Upsilon_V^{-1}(\mu)
    =\bm p_V^{-1}\!\left((\mu(B_i))_{i=1}^4\right),
  \qquad \mu\in\Upsilon_V(\Theta).
\end{equation*}
Evaluation on each \(B_i\) is continuous in TV, and \(\bm p_V^{-1}\) is continuous. Hence \(\Upsilon_V^{-1}\) is continuous and maps compact subsets of \(\Upsilon_V(\Theta)\) to compact subsets of \(\Theta\), establishing the properness criterion in Theorem~\ref{thm:global_identifiability}.
\end{proof}

Since the proof establishes injectivity of \(\bm p_V\), Corollary~\ref{cor:minimum_thresholds_global_identifiability} also gives global mean identifiability on \(\Theta\) with respect to \(\bm f(\nu)=(\ind_{B_i}(\nu))_{i=1}^4\) under \(\mathfrak S_V\) for every sufficiently small positive \(V\).

Both corollaries establish the existence of a sufficiently small positive threshold range. An explicit bound on \(V_*\) requires further quantitative estimates.

\section{Numerical Experiments}
\label{sec:numerical_experiments}

This section examines the probabilities constructed in Section~\ref{sec:poisson_identifiability}, from their sensitivity to their use in estimation. We study convergence of the finite-threshold Jacobian to its low-threshold limit, examine Jacobian ranks and probability-vector separation across the parameter domain, and evaluate finite-sample estimation from five categories of the first recorded crossing time.

\begin{figure*}[!t]
  \centering
  \includegraphics[width=0.90\textwidth]{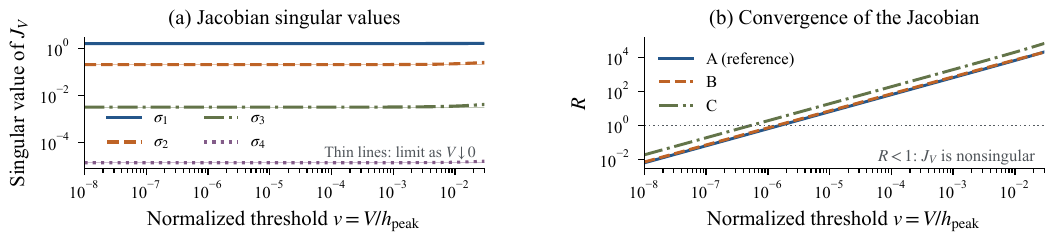}
  \caption{Local identifiability and parameter sensitivity using the fine timing setting and normalized parameter coordinates. (a) Singular values of \(J_V=D_{\bm z}\bm p_V\) at \(\bm\theta_{\mathrm A}=(120,0.15,0.30,1.50)\); thin lines show their low-threshold limits. (b) Relative Jacobian perturbation \(R=\|J_V-J_{\mathrm{lim}}\|_2/\sigma_{\min}(J_{\mathrm{lim}})\), at A, \(\bm\theta_{\mathrm B}=(90,0.12,0.25,1.25)\), and \(\bm\theta_{\mathrm C}=(180,0.18,0.35,1.75)\), with \(J_{\mathrm{lim}}=D_{\bm z}\bm p_{\mathrm{lim}}\). The dotted line marks \(R=1\); \(R<1\) guarantees that \(J_V\) is nonsingular.}
  \label{fig:identifiability_local}
\end{figure*}

\begin{figure*}[!t]
  \centering
  \includegraphics[width=0.90\textwidth]{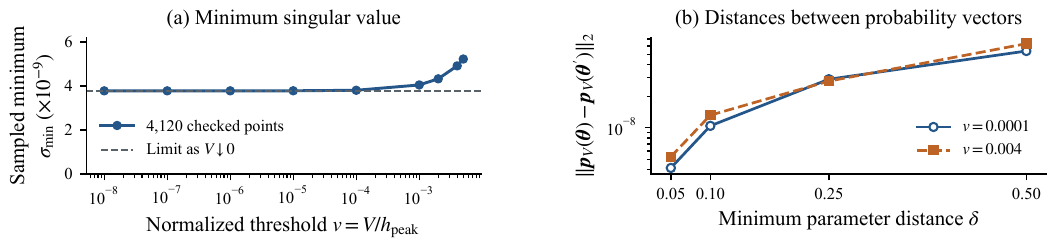}
  \caption{Global numerical checks on \(\overline{\Theta}\) using the fine timing setting. (a) Minimum of \(\sigma_{\min}(J_V)\) over 4,096 Sobol points, 16 corners, and eight face centers at each threshold; the dashed line shows the minimum of \(\sigma_{\min}(J_{\mathrm{lim}})\) over the same points. (b) Smallest feasible probability-vector distance found from 12 initial guesses under \(\|\bm z-\bm z'\|_2\geq\delta\). Each reported distance is an upper bound on the corresponding global minimum.}
  \label{fig:identifiability_global}
\end{figure*}

\begin{figure*}[!t]
  \centering
  \includegraphics[width=\textwidth]{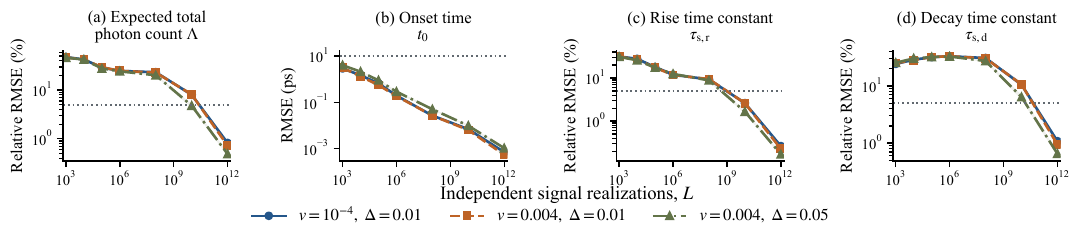}
  \caption{Parameter-estimation root-mean-square error (RMSE) versus the number of independent signal realizations \(L\), using five categories of the first recorded crossing time at \(\bm\theta_{\mathrm A}=(120,0.15,0.30,1.50)\). Each point summarizes 200 independently generated multinomial data sets with total count \(L\). Shading: pointwise 95\% percentile-bootstrap confidence intervals for RMSE from 1,000 resamples~\cite{efron1993introduction}. Dotted lines: 5\% relative-RMSE and 10 ps onset-time RMSE targets, with \(\tau_{\mathrm r}=1\) ns. The configuration with larger \(\Delta\) also changes the recorded-time cutoffs.}
  \label{fig:identifiability_recovery}
\end{figure*}

\subsection{Experimental Setup and Numerical Validation}

We use \eqref{eq:scintillation_example_model}, taking \(A\) and \(\tau_{\mathrm r}\) as voltage and time units, respectively. In these units, \(A=\tau_{\mathrm r}=1\), \(\tau_{\mathrm d}=4\), the observation window is \([0,2]\), and \(\bm\theta=(\Lambda,t_0,\tau_{\mathrm{s,r}},\tau_{\mathrm{s,d}})\) lies in
\begin{equation}
  \Theta=(80,200)\times(0.10,0.20)\times(0.20,0.40)\times(1,2).
  \label{eq:numerical_parameter_region}
\end{equation}
Scans and constrained fits include the boundary of this parameter set and therefore use \(\overline{\Theta}\). We use \(\bm\theta_{\mathrm A}=(120,0.15,0.30,1.50)\), denoted by A, as the reference parameter vector. Each coordinate \(z_j\) is obtained by subtracting the lower parameter bound and dividing by the parameter range; derivatives and parameter distances use these normalized coordinates. Thresholds are reported as \(v=V/h_{\mathrm{peak}}\), where \(h_{\mathrm{peak}}=\max_t h(t)\).

We use two timing settings, each specified by a quantization step \(\Delta\) and recorded-time cutoffs \(d_i=\Delta\kappa_i\). The fine setting uses \(\Delta=0.01\) and \(\bm d=(0.20,0.22,0.25,0.30)\); the coarse setting uses \(\Delta=0.05\) and \(\bm d=(0.20,0.25,0.30,0.35)\). Their comparison therefore reflects the combined effect of the quantization step and the cutoffs.

The analog-time cutoffs are \(\bm c=\bm d+(\Delta/2)\bm1\). The first cutoff is 0.205 in the fine setting and 0.225 in the coarse setting, both later than the largest onset, 0.20, in \(\overline{\Theta}\). Together with \eqref{eq:numerical_parameter_region}, these choices satisfy the parameter-domain and cutoff conditions of Corollary~\ref{cor:minimum_thresholds_global_identifiability}. All cutoffs precede the peaks of the responses to photons arriving within the model's support, so the signal is nondecreasing over the relevant interval. Thus the probabilities from Section~\ref{sec:poisson_local_identifiability} satisfy \(p_i(V,\bm\theta)=\Prob\{Y_{\bm\theta}(c_i)>V\}\). Writing \(p_i=p_i(V,\bm\theta)\), the four nested cutoff conditions give five mutually exclusive categories with probabilities
\begin{equation}
  \bm q_V=(p_1,p_2-p_1,p_3-p_2,p_4-p_3,1-p_4).
  \label{eq:numerical_five_categories}
\end{equation}
Here \(p_i=F_{V,\bm\theta}(d_i)\) are values of the unconditional first recorded crossing-time CDF. The last category includes later first recorded crossings or no recorded crossings. Writing \(q_j\) for the components of \(\bm q_V\), the inverse relation \(p_i=\sum_{j=1}^{i}q_j\), \(i=1,\ldots,4\), shows that \(\bm p_V\) and \(\bm q_V\) determine one another. Thus the mean-identifiability results for the four event indicators in Section~\ref{sec:poisson_identifiability}, in the sense of Definition~\ref{def:mean_identifiability}, also establish identifiability from the categorical distribution under the same conditions.

Each signal realization contributes one category. For \(L\) independent realizations sharing a common \(\bm\theta\), the category counts follow a multinomial distribution with probabilities \(\bm q_V\). These counts are the observations used in the estimation experiments below.

We compute finite-threshold probabilities using an amplitude power-series expansion of the compound-Poisson CDF through degree 12, as detailed in Appendix~\ref{app:numerical_evaluation}. The degree refers to the threshold amplitude, rather than a truncation of the photon count. Parameter derivatives use complex steps~\cite{martins2003complex} of size \(10^{-25}\) in the normalized coordinates. Checks using higher expansion degrees, 70-digit arithmetic, and independent Fourier inversion~\cite{gil1951note} gave maximum probability and Jacobian discrepancies of \(7.67\times10^{-15}\) and \(2.25\times10^{-15}\), respectively. At the checked locations, these are well below the probability separations and smallest singular values discussed below. An independent simulation of three million photon-arrival realizations with time quantization gave a maximum CDF discrepancy of \(9.47\times10^{-4}\), within the 95\% simultaneous Monte Carlo band \(1.72\times10^{-3}\)~\cite{massart1990tight}.

\subsection{Identifiability and Parameter Sensitivity}

The following checks use the fine timing setting. Let \(J_V=D_{\bm z}\bm p_V\), \(J_{\mathrm{lim}}=D_{\bm z}\bm p_{\mathrm{lim}}\), and \(R=\|J_V-J_{\mathrm{lim}}\|_2/\sigma_{\min}(J_{\mathrm{lim}})\). The ratio \(R\) measures the Jacobian perturbation relative to the smallest singular value of the limiting Jacobian. The condition \(R<1\) guarantees nonsingularity of \(J_V\); \(R\geq1\) gives no conclusion from this sufficient condition.

Figure~\ref{fig:identifiability_local} shows convergence to the limiting Jacobian, with \(R<1\) at sufficiently low tested thresholds. At A, all four computed singular values remain positive across the tested range and span about five orders of magnitude in the normalized parameter coordinates. This indicates local distinguishability together with substantial sensitivity differences among parameter directions.

Figure~\ref{fig:identifiability_global} extends the checks across \(\overline{\Theta}\). All sampled Jacobians have full rank, with the smallest sampled singular values of order \(10^{-9}\). As \(V\downarrow0\), these sampled minima approach the sampled minimum for \(J_{\mathrm{lim}}\). The parameter-pair searches find well-separated parameters with very similar first recorded crossing-time CDF values. Each reported feasible distance is an upper bound on the corresponding global minimum. Three searches producing the selected distances reached the iteration limit. Separately, at \(v=10^{-4}\), 64 noiseless probability vectors were each inverted from 32 initial guesses, with maximum normalized component error \(1.28\times10^{-9}\). These scans, searches, and inversions provide numerical evidence at the tested locations and reveal weak parameter sensitivity.

\subsection{Estimation from Categorized Crossing Times}

For Fig.~\ref{fig:identifiability_recovery}, we consider three estimation configurations: \((v,\Delta)=(10^{-4},0.01)\), \((0.004,0.01)\), and \((0.004,0.05)\), using the cutoff vector associated with each timing setting. We generate 200 independent multinomial data sets at A for each configuration and each total count \(L\in\{10^3,10^4,10^5,10^6,10^8,10^{10},10^{12}\}\), using the probabilities in \eqref{eq:numerical_five_categories}. These are finite-threshold numerical experiments; neither \(v=10^{-4}\) nor \(v=0.004\) is certified to lie in the global threshold range by an explicit bound on \(V_*\). We estimate the parameters by numerically maximizing the multinomial likelihood over \(\overline{\Theta}\). We minimize the equivalent multinomial deviance using bounded nonlinear least squares, initialized at \(\bm z_{\mathrm A}\) and \((0.8,0.2,0.2,0.8)\), and retain the fit with the smaller objective. The implementation adds starts and increases the iteration budget if the selected fit fails to converge; no such retries were needed for the reported data sets. Numerical termination does not certify a global likelihood optimum. All estimates are retained; an estimate is classified as a boundary estimate if any normalized coordinate is within \(10^{-6}\) of a bound.

At A, \(v=0.004\), \(\Delta=0.01\), and \(L=10^6\), the onset-time RMSE is 0.187 ps when \(\tau_{\mathrm r}=1\) ns. The relative RMSEs of \(\Lambda\), \(\tau_{\mathrm{s,r}}\), and \(\tau_{\mathrm{s,d}}\) are approximately 25\%, 12\%, and 33\%, with 98\% boundary estimates. Thus accurate onset estimation coexists with substantial uncertainty in the other three parameters. The onset-time RMSE concerns the common onset estimated from all \(L\) realizations and does not represent single-realization timing resolution. At \(L=10^{12}\), the three relative RMSEs fall below 1\% and the boundary rate is zero.

Across all three configurations in Fig.~\ref{fig:identifiability_recovery}, \(L=10^{12}\) is the first tested sample size meeting all four accuracy targets. The experiment does not determine the exact minimum sample size required to meet these targets.

At A, 2.62 of the 120 expected photons arrive before the last fine cutoff. The five categories therefore summarize early crossing-time behavior, whose weak sensitivity to some parameter changes is consistent with the substantial finite-sample errors.

\section{Discussion}
\label{sec:discussion}

The theoretical and numerical results separate the threshold requirement for parameter uniqueness from the sensitivity needed for accurate estimation. We discuss the resulting design choices, the information retained by the selected observations, and extensions of the analysis.

\subsection{Implications for Identifiability and Threshold Requirements}

A threshold defines a crossing-time distribution with multiple parameter-dependent probabilities. The construction in Corollaries~\ref{cor:minimum_thresholds_local_identifiability} and~\ref{cor:minimum_thresholds_global_identifiability} shows that the number of thresholds need not match the number of unknown parameters.

The rank inequalities in Proposition~\ref{prop:derivatives_through_sampling} can identify stages at which first-order parameter distinctions are lost through crossing, timing, and selection. Attaining the minimum threshold count leaves open the choice of threshold levels, time quantization, and recorded statistics for accurate estimation. These design choices should therefore be assessed using sensitivity and estimation error as well as identifiability.

\subsection{Identifiability and Parameter Estimation}

Identifiability establishes parameter uniqueness, while estimation accuracy also depends on the sensitivity and statistical variability of the observations, the sample size, and the estimator~\cite{raue2009structural}. The small singular values and probability-vector separations in Figs.~\ref{fig:identifiability_local} and~\ref{fig:identifiability_global} reveal weak sensitivity in some parameter directions. Such directions can amplify the effect of sampling fluctuations on parameter estimates, consistent with the errors and boundary estimates in Fig.~\ref{fig:identifiability_recovery}.

The four cutoff events in the single-threshold proofs were chosen to establish uniqueness. Equation~\eqref{eq:numerical_five_categories} re-expresses their probabilities invertibly. The reduction from all recorded crossings in each realization to one of five categories has not been shown to preserve all information relevant to estimation. Other cutoff choices and a broader set of recorded crossings may improve sensitivity and estimation accuracy.

\subsection{Scope and Further Research}

The general results concern a fixed sampler and rely on the regularity and rank assumptions of Theorems~\ref{thm:local_identifiability} and~\ref{thm:global_identifiability}; the global criterion additionally requires the stated image geometry. The internal randomness used by crossing, timing, and selection is mutually independent and independent of the signal; the stage outputs remain linked by sequential sampling.

The single-threshold corollaries concern the specified four-parameter photon-arrival intensity, a fixed known single-photon response, and deterministic sampling, with their respective cutoff and parameter-domain conditions. They establish the existence of a sufficiently small positive threshold range. A first direction is to obtain explicit bounds on this range and to quantify how threshold levels and cutoff choices affect sensitivity.

A second direction is to use joint distributions of recorded crossings to analyze error propagation and bias and to compare estimator performance. With a specified prior and loss function, this includes comparisons of Bayes risk against the minimum Bayes risk for the chosen observations. Such analysis could guide weight optimization for amplitude-domain and time-domain weighted least squares and comparisons between the resulting estimators.

A further direction is joint identification of response, sampler, and intensity parameters, which requires re-examining identifiability for the enlarged parameterization. At the system level, these analyses can support joint design of threshold settings, time quantization, and estimators, with estimation gains weighed against implementation costs.

\section{Conclusion}
\label{sec:conclusion}

We established conditions for local and global parameter identifiability from distributions of recorded crossings under a fixed MT sampler. The ranks of distribution derivatives cannot increase through crossing, timing, and selection. Under the respective regularity, rank, and topological assumptions, local identifiability is characterized by a nonsingular Jacobian of finitely many event probabilities, and global identifiability by properness of the parameter-to-distribution map onto its image.

For the specified nonhomogeneous Poisson signal model with four intensity parameters, a fixed known single-photon response, and deterministic MT sampling with time quantization, a single sufficiently small positive threshold ensures local and global identifiability under their respective conditions. The construction uses four cumulative probabilities of the first recorded crossing time and attains the minimum number of thresholds for this model and sampler. Numerical experiments reveal weak sensitivity and substantial finite-sample errors when estimating shared parameters from the corresponding five categories. The minimum threshold count establishes uniqueness under the stated conditions, while accurate estimation also requires attention to parameter sensitivity.

\appendices
\section{Finite-Threshold Probability Evaluation}
\label{app:numerical_evaluation}

This appendix describes the evaluator used in Section~\ref{sec:numerical_experiments}. Fix an analog-time cutoff \(c\) and parameters \(\bm\theta\), and let \(a(x)\) be the inverse of the increasing branch of the response \(h\). The cutoff interval precedes the response peak. For \(0<V<h(c-t_0)\), define
\[
  \begin{aligned}
    \mu_c&=\int_{t_0}^{c}\lambda_{\bm\theta}(u)\,du,\\
    \nu_c(x)&=\lambda_{\bm\theta}(c-a(x))a'(x),\quad 0\le x\le V.
  \end{aligned}
\]
Here \(\nu_c\) is the intensity density of individual photon contributions at \(c\), restricted to amplitudes at most \(V\). The full photon count before \(c\) has mean \(\mu_c\), so contributions larger than \(V\) remain accounted for by the Poisson factor below.

Expand the transformed intensity about zero as
\[
  V\nu_c(Vy)=\sum_{j\ge0}w_jy^j.
\]
For sufficiently small positive \(V\), expansion of the Poisson convolutions and integration of their monomials over the amplitude simplex give
\[
  f_0=1,\qquad
  f_n=\frac1n\sum_{j=1}^{n}
       \frac{w_{j-1}f_{n-j}}{\binom{n}{j}},\quad n\ge1,
\]
and
\[
  \Prob\{Y_{\bm\theta}(c)\le V\}
      =e^{-\mu_c}\sum_{n\ge0}f_n.
\]
Indeed, each monomial integration contributes the corresponding factorial ratio, and collecting terms of total amplitude degree \(n\) yields the displayed recurrence. Since \(w_j\) is proportional to \(V^{j+1}\), \(f_n\) has degree \(n\) in \(V\). The term \(f_0\) includes the zero-photon atom. In particular, the approximation through degree \(m\) is evaluated as
\[
  p_i(V,\bm\theta)\approx
  -\operatorname{expm1}(-\mu_{c_i})
  -e^{-\mu_{c_i}}\sum_{n=1}^{m}f_n,
\]
with the coefficients evaluated at \(c_i\). This form avoids subtracting a CDF close to one.

The coefficients of \(a\) are obtained by Lagrange inversion,
\[
  [x^n]a(x)=\frac1n[t^{n-1}]\left(\frac{t}{h(t)}\right)^n,
\]
where brackets denote coefficient extraction. Composition with the exponential intensity then gives \(w_j\). The default degree is \(m=12\); refinement uses degree 20 and an independent degree-28 calculation with 70-digit arithmetic. The evaluations for Figs.~\ref{fig:identifiability_local}--\ref{fig:identifiability_recovery} satisfy \(0<V<h(c_i-t_0)\) at every cutoff.

The independent derivative check uses 70-digit central differences with step \(10^{-18}\) in normalized coordinates. Fourier inversion splits photons according to whether their age at \(c\) exceeds \(a(V)\). An older photon alone exceeds the threshold. The younger contributions are scaled by \(V\); their zero- and one-photon terms cannot exceed the threshold and are removed analytically before inversion. Frequency cutoffs of 256, 512, and 1,024 and steps of 0.25 and 0.125 provide refinement checks. These comparisons assess numerical accuracy at the checked locations and do not supply a uniform error certificate over the parameter domain.

\noindent\begin{minipage}{\columnwidth}
\vspace{\baselineskip}
\section*{Acknowledgment}

The authors used OpenAI Codex to assist with drafting and revising the abstract, Sections~\ref{sec:introduction}--\ref{sec:conclusion}, and the appendix, checking mathematical derivations, and developing the numerical code used in Section~\ref{sec:numerical_experiments}. The authors take full responsibility for the final content.
\end{minipage}

\IEEEtriggeratref{8}
\bibliographystyle{IEEEtran}
\bibliography{MT_identifiability-ref}

\end{document}